 \documentclass{amsproc}

\usepackage{eurosym}
\usepackage{amsmath}
\usepackage{graphicx}
\usepackage{url}

\numberwithin{equation}{section}
\newtheorem{theorem}{Theorem}

\newtheorem{lemma}{Lemma}

\newtheorem{proposition}{Proposition}

\input{tcilatex}

\begin{document}
\title{On stochastic realism and CP bias in diffractive dissociations}
\author{A.Y.Klimenko}
\address{CMES, SoMME, The University of Queensland, Brisbane, Australia}
\email{a.klimenko@uq.edu.au}
\thanks{}
\date{}

\begin{abstract}
CP bias in proton--antiproton diffractive dissociation can be viewed from
several philosophical perspectives, notably stochastic realism and
stochastic epistemicism. When combined with the presumption of temporal
symmetry, stochastic realism suggests that the laws of physics allow the
possibility of thermodynamic antisystems, although whether antisystemic
characteristics can be physically realised in nature remains an open
question.

Here, this bias is interpreted as apparent, indicating significant
non-unitary contributions and a distinction from fundamental CP violations
arising from unitary Hamiltonian dynamics. Such apparent effects may arise
from intrinsic stochasticity, from environmental interactions, or from
interference between the two. This work investigates the relevant mechanisms
and determines conditions for creating, transmitting, or screening a CP bias.

In the environmental branch, an equilibrated radiation bath may transmit a
CP bias from matter-dominated surroundings, although causality constraints
may limit this possibility. In the intrinsic branch, the observed bias is
consistent with subleading antisystemic effects required by CPT invariance.
Further experiments are needed to distinguish between these mechanisms.
\end{abstract}

\maketitle
\tableofcontents









\bigskip

\section{Introduction}

The analysis of proton--antiproton and proton--proton diffractive
dissociation indicates the presence of a statistically significant and
consistent bias ($\approx $20\% if expressed as the cross-section ratio and $%
\approx $10\% as the dephasing rate anomaly \cite{Klimenko2026CDLindblad}).
This bias requires a plausible explanation, preferably one that can be
tested experimentally. A large, consistent experimental error and a
fundamental CP violation of the unitary Hamiltonians are both possible in
principle but unlikely: the former because it would have to act coherently
across independent experiments, the latter because a CP violation of this
magnitude in the strong sector is implausible.

The interpretation adopted here is that the observed bias is not
fundamental, but apparent. That is, it is not attributed to CP-violating
terms in the unitary Hamiltonian, but to non-unitary interference effects
that are not captured by a purely unitary description. Such effects are
naturally associated with decoherence and emergence of the arrow of time as
the outgoing hadronised fragments must lose phase coherence to form
separated final states. This requires an appropriate framework that can
characterise quantum symmetries in the context of non-unitary evolution.

Two possible sources of these non-unitary effects are considered:
environmental interactions and intrinsic stochastic disturbances. This work
implements a new framework for analysing fundamental symmetries in
non-unitary processes to conceptually examine both possibilities. While the
intrinsic branch has, to a large extent, been introduced previously \cite%
{Milburn1991IntrinsicDecoherence,Schulman1997TimesArrowsBook,Beretta2015,ScharnhorstWolpertRovelli2024BoltzmannBridges,Klimenko2026CDLindblad}%
, the environmental branch \cite%
{Zurek2003DecoherenceEinselection,Breuer2007book,LindbladIntro2020} requires
further analysis, which is presented in the current work. Adjudicating
between these possibilities requires more than empirical input: it depends
on foundational assumptions about temporal structure and interpretation of
the physical states, which the following sections make explicit.

Sections~2--5 review the philosophical and physical principles relevant to
the analysis (e.g. stochastic realism and epistemicism, temporal symmetry,
arrow of time, antecedent causality, dual temporal boundary conditions and
antisystems). Section~6 presents the asymmetry as apparent rather than
fundamental CP violation. Section~7 analyses environmental interference,
while Section~8 considers interaction of intrinsic and environmental
mechanisms. The analysis is then summarised in the Discussion and Conclusion
sections.

\section{Stochastic realism versus stochastic epistemicism}

\begin{figure}[tbp]
\includegraphics[width=\textwidth,page=1,trim=0cm 0cm 0cm 0cm, clip
]{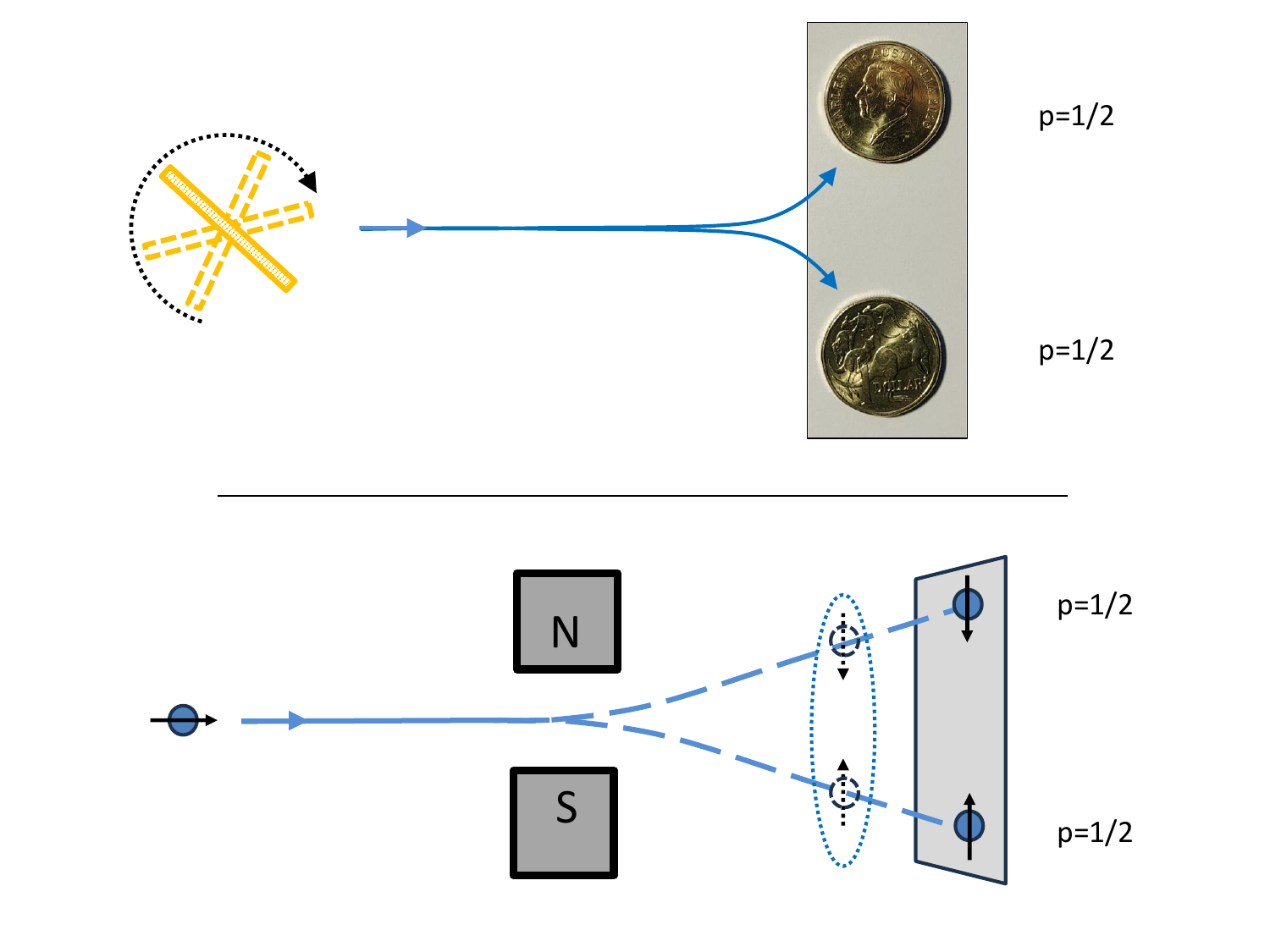}
\caption{Stochastic epistemicism illustrated by coin toss and stochastic
realism illustrated by wave function collapse}
\label{fig1}
\end{figure}

\emph{Randomness} is commonly viewed from an \emph{epistemic perspective} as
arising from a lack of knowledge about the system under consideration. This
lack of knowledge induces a degree of uncertainty in the outcome, and this
uncertainty is recognised as stochasticity. A standard example is a coin
toss, illustrated in Figure \ref{fig1}, top: if we knew the exact initial
conditions of the coin at the time of the toss, we would be able to predict
the outcome. Therefore, the apparent randomness of the outcome --- either
heads or tails --- is not necessarily a property of the real world itself,
but rather a consequence of our limited knowledge of it.

If the most extreme viewpoints are excluded, the epistemic perspective on
randomness does not generally claim that physical stochasticity does not
exist. Rather, it suggests that whether or not such stochasticity exists, it
does not need to be explicitly considered, because treating randomness as a
lack of knowledge about the relevant details is sufficient for most
practical purposes. In this sense, the epistemic perspective is more
concerned with explaining our observations than with making claims about
physical reality itself.

In certain formulations of quantum and statistical mechanics ---
particularly those involving open systems and decoherence \cite%
{Breuer2007book,LindbladIntro2020,Zurek2003DecoherenceEinselection} --- the
epistemic perspective is represented by environmental degrees of freedom
that are not exactly known. If we knew these, we would be able to predict
the state of the system exactly, but we do not and must resort to
coarse-graining over uncertain details --- we are not interested in the
specifics of the environmental interactions but rather in their aggregate
presence. This approach is called here \emph{stochastic epistemicism}, and
in many cases it is sufficient to explain our observations.

\bigskip

\bigskip

The alternative view --- referred to here as \emph{stochastic realism} ---
is based on an \emph{ontic} \emph{interpretation of randomness}. It holds
that stochasticity is real and substantially present in the universe. A
standard example illustrating stochastic realism is the measurement of a
quantum spin state prepared in a superposition, that collapses with 50\%
probability into either the spin-up or spin-down state, as illustrated in
Figure \ref{fig1}, bottom. According to our current physical understanding,
no amount of prior information would allow us to predict the individual
outcome with certainty; this is therefore taken to illustrate genuine, or
ontic, randomness. At the elementary quantum level, the increase of
randomness is associated with decoherence. In stochastic realism, this
decoherence is not merely an epistemic effect produced by tracing out
inaccessible degrees of freedom, but is induced by an intrinsic mechanism
involving inherent stochastic variations \cite%
{Milburn1991IntrinsicDecoherence,Beretta2009,Beretta2015}.

According to stochastic realism, the observed increase of entropy is a
physical process, characterising an increasing amount of randomness (or
disorder) in the universe. By contrast, stochastic epistemicism would regard
this increase only as an apparent effect associated with coarse-graining:
that is, with the way we observe, describe and analyse our surroundings.

Stochastic realism does not deny the existence of epistemic uncertainty.
Rather, it states that physical randomness is also present, and that it is a
substantial feature of reality.

\begin{proposition}
Stochastic realism and stochastic epistemicism are philosophical
perspectives rather than directly testable physical theories. The
perspectives themselves cannot be tested experimentally; but a physical
theory framed within, or naturally associated with, one of them --- usually
with some freedom of interpretation --- can be tested.
\end{proposition}

This proposition can be illustrated by example. If the randomness is taken
to be intrinsic to the microscopic system, entering as stochastic terms in
its governing equations, this corresponds to stochastic realism. If the same
randomness is instead attributed to the particle's interactions with a
presumed omnipresent background field, it corresponds to stochastic
epistemicism --- even if the two interpretations are physically
indistinguishable in their observable predictions.

\section{Temporal symmetry and emergence of the time arrow}

Although everyday experience presents the past and the future as radically
different, the fundamental dynamical laws of physics --- in their classical,
relativistic, and quantum forms --- are largely \emph{time-reversible} and
nearly \emph{time-symmetric}, apart from qualifications such as
weak-interaction time-reversal violation \cite%
{Symmetry1972,PDGConservation2020}.

The standard explanation of this tension appeals to the \emph{Past Hypothesis%
} \cite{Albert2000TimeChance}, according to which the universe originated in
an exceptionally special, \emph{low-entropy state}. The central question is
not whether such a condition is needed (this is commonly recognised), but
whether it is sufficient by itself. Reichenbach's \emph{principle of the
parallelism of entropy increase} \cite{Reichenbach1956Direction} suggests
that an adequate explanation must also account for why entropy gradients in
quasi-isolated branch systems are aligned. Thus the Past Hypothesis alone
may leave open the question of why (and maybe if) a local arrow of time
reappears or persists after an isolated system has spent a long time in
equilibrium.

Stochastic realism provides a possible answer. On this view, stochastic
processes are constrained by initial conditions in the remote past, with no
relevant low-entropy final condition in the accessible future. Their
evolution is therefore forward-directed: it is described by stochastic
semigroups rather than reversible dynamical groups. For example, if
stochastic particles reach a uniform distribution in a confined region, that
distribution is stationary only relative to the existing confinement. If the
available region expands, the old uniform state is no longer an equilibrium
state of the enlarged system, and ordinary forward diffusion resumes. Thus,
on the stochastic-realist picture, prolonged equilibrium does not erase the
arrow of time: once the constraints change, the forward stochastic law
continues to determine the direction of relaxation.

In this picture, deterministic components of the governing equations remain
time-symmetric while the thermodynamic arrow of time is primed and selected
by stochastic components conditioned by the Past Hypothesis. By contrast, in
purely Hamiltonian or unitary dynamics, the evolution is reversible. After
an isolated system has remained at equilibrium for a long time, the Past
Hypothesis alone may not determine the local continuation of the time arrow
without additional assumptions such as Reichenbach's branch independence.

\bigskip

\bigskip

\begin{figure}[tbp]
\includegraphics[width=\textwidth,page=2,trim=0cm 0cm 0cm 5cm, clip
]{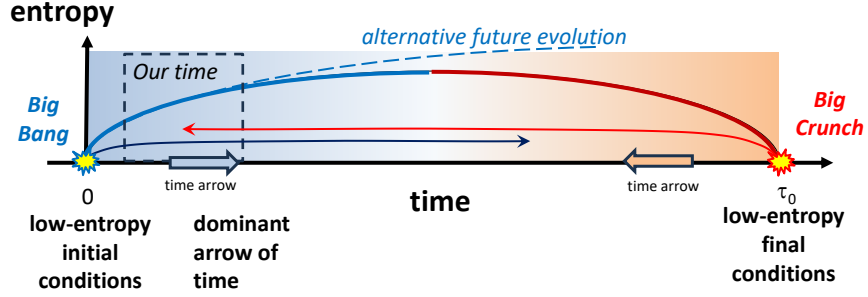}
\caption{Schematic of dual temporal boundary conditions imposed on the
universe. The dashed line shows an alternative evolution without final
conditions (or with more remote final conditions). Large arrows indicate
local time arrows, while lines with arrows show systems travelling into
``antiworld'' and antisystems emerging in our time window. }
\label{fig2}
\end{figure}

\bigskip

According to the Past Hypothesis, the \emph{arrow of time} is determined by
the very special, low-entropy initial conditions imposed on the Universe. On
the Penrose--Schulman-type view adopted here, this condition is an
independent restriction placed on the admissible histories of the Universe 
\cite%
{Penrose1979Singularities,Schulman1991TwoTime,Schulman1997TimesArrowsBook}.
This approach preserves the time-symmetric character of the underlying
physical laws and is the scientifically preferred framework for the present
discussion. An alternative is provided by Gold's picture \cite%
{Gold1962Arrow,Price1996TimesArrow}, in which low-entropy conditions are
tied to cosmological singularities rather than imposed as independent
restrictions. In that picture, the thermodynamic arrow is fixed by
cosmological evolution rather than by an independent boundary condition, and
the relevant physical laws acquire an effective temporal directionality
through their dependence on cosmological time. While Gold's approach is
noted here as a contrasting possibility, the present analysis follows the
well-established Past-Hypothesis framework. Despite their differences, both
approaches allow, at least in principle, for a future low-entropy temporal
boundary condition that might be attached to a Big Crunch, although for
different reasons. This picture is illustrated in Figure \ref{fig2}.

If the fundamental laws are genuinely time-symmetric and low-entropy
conditions are independent, then the laws cannot, by themselves, forbid the
time-reversed counterpart of the Past Hypothesis. Just as the laws permit a 
\emph{low-entropy boundary condition in the past}, they also permit a \emph{%
low-entropy boundary condition in the future}. Otherwise, the laws
themselves would already contain a built-in temporal asymmetry. Whether such
a future boundary condition is actually realised is not essential here:
within the time window considered (Figure \ref{fig2}), the universe exhibits
similar dynamics irrespective of the final condition. The important point is
that such a condition could exist and, if imposed, would produce \emph{%
another family of valid solutions} of the same physical laws. These
solutions are labelled here with the prefix `\emph{anti-}'. In stochastic
formulations, initial and final constraints can also be applied
simultaneously (as in Figure \ref{fig2}), giving rise to bridge-type
processes: stochastic evolutions conditioned both on a past temporal
boundary condition and on a future temporal boundary condition \cite%
{Schulman1991TwoTime,ScharnhorstWolpertRovelli2024BoltzmannBridges,MovillaMiangolarraSabbaghGeorgiou2025QuantumSchrodingerBridges}%
.

When considered \ from the perspective of stochastic realism, the
time-symmetric formulation of the laws of nature involves emergence of a
prominent \emph{time arrow} and, with it, \emph{antecedent causality}, in
the initial stages of the development shown by the blue background in Figure %
\ref{fig2}. While forward-diffusive systems do exist, and provide evidence
for a thermodynamic arrow associated with a low-entropy past state,
reverse-diffusive systems are not excluded by our current understanding of
the laws of physics. Whether or not final low-entropy conditions are
realised in our Universe, time-symmetric laws allow for such conditions and
therefore for corresponding \emph{antisystems} with reverse-time diffusion.
This does not mean that such antisystems have been observed; it means that
they are valid solutions permitted by the time-symmetric structure of the
underlying physical laws \cite{Klimenko2025TwoTypesTemporalSymmetry}. The
presence of antisystemic properties can be determined only by examining such
solutions in the context of the conventional systemic surroundings common in
our Universe. As shown in Figure \ref{fig2}, antisystemic solutions can, at
least in principle, exist in our time window: \ 

\begin{proposition}
Although thermodynamic antisystems appear to be allowed by the laws of
nature as currently understood, the central question is whether any evidence
of their existence can be found in the real world.
\end{proposition}

The formalism required to investigate these antisystemic solutions is
developed in the next section.

\section{Equations governing stochastic systems with dual temporal conditions%
}

As discussed in the previous sections, stochastic realism, together with the
expectation of temporal symmetry in dynamical equations, naturally leads to
a \emph{stochastic formulation} with dual temporal boundary conditioning for 
\emph{general dynamical systems} \cite%
{ScharnhorstWolpertRovelli2024BoltzmannBridges,Klimenko2025TwoTypesTemporalSymmetry}%
: 
\begin{equation}
dx^{i}=a^{i}(\mathbf{x},t)\,dt+\sum_{j}b^{ij}(\mathbf{x},t)\circ
dw^{j},\qquad \mathbf{x}(t_{1})=\mathbf{x}_{1},\qquad \mathbf{x}(t_{2})=%
\mathbf{x}_{2}.  \label{sde-x}
\end{equation}%
For \emph{quantum-mechanical systems}, the corresponding stochastic equation
may be written as \cite%
{GisinPercival1992QSDOpen,WisemanMilburn2010QuantumMeasurementControl,DephasingHAM2021,StochasticHamiltonians2026,Klimenko2026CDLindblad}
\begin{equation}
i\hbar \,d\rho =\left[ H,\rho \right] \,dt+\sum_{j}\left[ L_{j},\rho \right]
\circ dw^{j},\qquad \rho (t_{1})=\rho _{1},\qquad \rho (t_{2})=\rho _{2}.
\label{sde-rho}
\end{equation}%
Here, $\rho $ is the density matrix\ and $w^{j}=w^{j}(t)$ denotes a set of
Wiener processes with zero mean $\left\langle dw^{j}\right\rangle =0$ and
correlation $\left\langle dw^{j}dw^{\ell }\right\rangle =\gamma ^{j\ell }dt$
where $\gamma ^{j\ell }$ is real. Temporal boundary conditions at $t=t_{1}$
and $t=t_{2}$ are assumed compatible. Equations (\ref{sde-x}) and (\ref%
{sde-rho}) reflect Kubo's concept of stochastic Liouville equations \cite%
{Kubo1963}. Equation (\ref{sde-rho}) can be converted into a system of real
equations and considered as a special case of (\ref{sde-x}). Note that
equation (\ref{sde-rho}) is generally different from standard stochastic Schr%
\"{o}dinger equations \cite{BieleDAgosta2012}, which contain drift-type
dephasing terms. A combination of random and drift dephasing terms generally
cannot be reversed in time since random and deterministic terms are
transformed differently. For this reason, such stochastic Schr\"{o}dinger
forms are not suitable for the time-symmetric formulation used here.

\emph{Dual-boundary stochastic formulations} are required to express
temporal symmetry in systems containing stochastic terms. In general, the
solution obtained by evolving stochastic trajectories forward in time from $%
t=t_{1}$ while imposing the additional boundary condition at $t=t_{2}$ need
not coincide with the solution obtained by evolving trajectories backward in
time from $t=t_{2}$ while imposing the additional boundary condition at $%
t=t_{1}$. In such cases, the two temporal directions are not equivalent, and
one must effectively nominate a preferred direction of time.

To obtain a time-symmetric formulation, the stochastic differential
equations must be understood in the \emph{Stratonovich sense} rather than in
the \emph{It\^{o} sense}, since the It\^{o} formulation is inherently
time-directional \cite{Klimenko2025TwoTypesTemporalSymmetry}. The
Stratonovich interpretation is indicated in equations (\ref{sde-x})--(\ref%
{sde-rho}) by the symbol \textquotedblleft $\circ $\textquotedblright . Only
under Stratonovich interpretation can $a^{i}$\ be understood as physical
quantities governing deterministic dynamics and $H$ be treated as the
physical Hamiltonian. However, the use of the Stratonovich interpretation
alone does not guarantee equivalence between the forward-time and
reverse-time formulations. Such equivalence is achieved under conditions
referred to here as \emph{odd-symmetric conditions} \cite%
{Klimenko2025TwoTypesTemporalSymmetry}. For general dynamical systems, these
conditions require the absence of divergence, 
\begin{equation}
\sum_{i}\frac{\partial a^{i}}{\partial x^{i}}=0,\qquad \sum_{i}\frac{%
\partial b^{ij}}{\partial x^{i}}=0.  \label{FP-cond}
\end{equation}%
For quantum systems, the corresponding requirement is that the operators be
Hermitian: 
\begin{equation}
H=H^{\dagger },\qquad L_{j}=L_{j}^{\dagger }.  \label{Lin-cond}
\end{equation}%
Under these conditions, entropy is not changed by the deterministic
components of the model, and the forward-time and reverse-time formulations
become equivalent. General Hamiltonian dynamics, as well as unitary
evolution in quantum mechanics, satisfy these conditions and preserve
entropy. Under odd-symmetric conditions (\ref{FP-cond})--(\ref{Lin-cond}),
the fundamental laws in the form of Eqs.~(\ref{sde-x})--(\ref{sde-rho}) can
be equivalently interpreted forward and backward in time, even with the
stochastic terms present.

There is also another class of temporal-symmetry conditions for stochastic
equations, referred to here as \emph{even-symmetric conditions}. These
conditions are based on Kolmogorov's principle of detailed balance \cite%
{Klimenko2025TwoTypesTemporalSymmetry} and are, in general, incompatible
with the odd-symmetric constraints discussed above (\ref{FP-cond})-(\ref%
{Lin-cond}). From a physical perspective, the odd-symmetric formulation is
consistent with conventional thermodynamic behaviour, whereas the
even-symmetric formulation may become relevant to thermodynamic behaviour in
the presence of strong gravitational fields. In the present work, only the
odd-symmetric case is considered.

\bigskip

The probability distribution function is factorised as $f(\mathbf{x}%
,t)=\varphi (\mathbf{x},t)\psi (\mathbf{x},t)/C,$ where $C$ is a
normalisation constant. The functions $\varphi (\mathbf{x},t)$ and $\psi (%
\mathbf{x},t)$ represent, respectively, the \emph{forward-time
entropy-increasing diffusion} and the \emph{reverse-time entropy-decreasing
diffusion}. They are subject to the boundary conditions 
\begin{equation}
\varphi (\mathbf{x},t_{1})=\delta (\mathbf{x}-\mathbf{x}_{1}),\qquad \psi (%
\mathbf{x},t_{2})=\delta (\mathbf{x}-\mathbf{x}_{2}).
\end{equation}%
The probability distribution then satisfies the equation 
\begin{equation}
\frac{\partial (\varphi \psi )}{\partial t}+\sum_{i}\frac{\partial
(a^{i}\varphi \psi )}{\partial x^{i}}=\sum_{i,j}\frac{\partial }{\partial
x^{i}}\left[ B^{ij}\left( \psi \frac{\partial \varphi }{\partial x^{j}}%
-\varphi \frac{\partial \psi }{\partial x^{j}}\right) \right] ,  \label{FP-f}
\end{equation}%
which combines the forward-time and reverse-time Fokker--Planck equations.
Here the symmetric diffusion matrix is defined as $%
B^{ij}=B^{ji}=b^{ik}b^{jk}/2.$ It is useful to distinguish three different
meanings of dual parabolicity---the presence of both forward-parabolic and
backward-parabolic terms---as discussed in Appendix \ref{app:CMC}.

The quantum analogue of this equation is the\emph{\ Lindblad dephasing
equation} \cite%
{Lindblad1976,GoriniKossakowskiSudarshan1976,Breuer2007book,LindbladIntro2020}
\begin{equation}
i\hbar \frac{d\tilde{\rho}}{dt}=\left[ H,\tilde{\rho}\right] -\sum_{j,\ell }%
\frac{i}{2\hbar }\gamma _{j\ell }\left[ L_{j},\left[ L_{\ell },\tilde{\rho}%
\right] \right] =\left[ H,\tilde{\rho}\right] -\sum_{j}\frac{i}{2\hbar }%
\gamma _{j}^{\circ }\left[ L_{j}^{\circ },\left[ L_{j}^{\circ },\tilde{\rho}%
\right] \right]  \label{eq:rho}
\end{equation}%
where $\tilde{\rho}$ denotes the average of the density matrix $\rho $ over
realisations of the stochastic processes $w^{j}(t)$. Since $\gamma _{j\ell }$
is real, the sum over $j$ and $\ell $\ can be diagonalised with a new
Hermitian set of operators $L_{j}^{\circ }=({L}_{j}^{\circ })^{\dagger }$
and without loss of generality as shown in (\ref{eq:rho}). In line with the
Fokker--Planck equation with dual parabolicity, the Lindblad dephasing
equation is generalised to represent both decoherence and recoherence \cite%
{Klimenko2026CDLindblad}. Forward-time evolution and associated \emph{%
decoherence} (loss of coherence) correspond to positive coefficients, $%
\gamma _{j}^{\circ }>0$, whereas reverse-time evolution and associated \emph{%
recoherence} (regaining coherence) correspond to negative coefficients, $%
\gamma _{j}^{\circ }<0$. The negative-coefficient case is not meant as an
ordinary forward-time Markovian Lindblad semigroup; it represents the
reverse-time or dual-boundary branch of the generalised dephasing
formulation used here \cite%
{Klimenko2016,SN-AS2021,Klimenko2025TwoTypesTemporalSymmetry,Klimenko2026CDLindblad}%
. The interval of validity of the generalised Lindblad equation is
restricted at both ends---in the past and in the future---to preserve
positivity of the density matrix. As the dephasing operator is unital (i.e.
preserves $\tilde{\rho}\sim I$), non-negative $\gamma _{j}^{\circ }\geq 0$
ensure that von Neumann entropy is non-decreasing in time \cite{Abe2017},
while\ negative $\gamma _{j}^{\circ }<0$ correspond to recoherence and to
entropy-decreasing behaviour. If all $\gamma _{j}^{\circ }=0$, this equation
reduces to unitary dynamics preserving entropy. The generalised Lindblad
dephasing equation (\ref{eq:rho}) combines all these possibilities. This
consideration is summarised by:

\begin{proposition}
The dual-parabolic Fokker--Planck equation in Hamiltonian mechanics and the
dual-signed Lindblad equation in quantum mechanics reflect stochastic
realism combined with an odd temporal symmetry, realised through dual
temporal boundary conditions.
\end{proposition}

Assuming that $H$ and $L_{j}^{\circ }$ are expandable in the same energy
basis 
\begin{equation}
\ H\left\vert n\right\rangle =E_{n}\left\vert n\right\rangle ,\ \ \text{ }%
L_{j}^{\circ }=\sum_{n}\lambda _{jn}\left\vert n\right\rangle \left\langle
n\right\vert  \label{eq_H_L}
\end{equation}%
\ \ \ complying with the pure dephasing condition $\left[ H,L_{j}^{\circ }%
\right] =0$ \cite%
{Albert2014ConservedLindblad,Fagnola2019MarkovianDephasing,Klimenko2026CDLindblad}%
, the system's density components $\tilde{\rho}_{nm}=\left\langle
n\right\vert \tilde{\rho}\left\vert m\right\rangle $ satisfy the equation 
\begin{equation}
\frac{d\tilde{\rho}_{nm}}{dt}=-\left( \frac{i}{\hbar }\Delta E_{nm}+\frac{%
\Gamma _{nm}^{S}}{\hbar ^{2}}\right) \tilde{\rho}_{nm}  \label{eq_Lind_nm}
\end{equation}%
where%
\begin{equation}
\Delta E_{nm}=E_{n}-E_{m},\ \ \ \Gamma _{nm}^{S}=\frac{1}{2}\sum_{j}\gamma
_{j}^{\circ }(\lambda _{jn}-\lambda _{jm})^{2}
\end{equation}

\section{Antisystems and the CP/CPT dichotomy in dephasing dynamics}

Do antisystems actually exist? This question remains open \cite%
{KlimenkoMaas2014,Klimenko2017KineticsCPT,Etesi2022,Klimenko2026Antisystems}%
. Although \emph{antisystems} ---that is, systems predominantly governed by
reverse-time evolution in (\ref{FP-f}) or (\ref{eq:rho})---are permitted by
the laws of physics \cite%
{Klimenko2025TwoTypesTemporalSymmetry,Klimenko2026Antisystems}, the
principal question is whether they, or at least some of their properties,
can be detected in the real world. Antimatter is the most natural known
candidate to consider in this context, although the possible connection
between antisystems and antimatter should not be assumed a priori. The
similarity between the terms \textquotedblleft antisystem\textquotedblright\
and \textquotedblleft antimatter\textquotedblright , therefore, may or may
not reflect the underlying physics.

Essentially, there are two principal incompatible possibilities for
non-unitary evolutions:

\begin{enumerate}
\item[\textbf{I.}] \textbf{CP invariance:} antimatter behaves as an ordinary
thermodynamic system.

\item[\textbf{II.}] \textbf{CPT invariance:} antimatter behaves as an
antisystem with reverse-time dynamics.
\end{enumerate}

The choice between possibilities I and II cannot be made without
experimental evidence. CP invariance, understood here as \emph{charge-parity
invariance}, implies that replacing matter with antimatter does not alter
the systemic properties of the system. By contrast, CPT invariance,
understood as \emph{charge-parity-time invariance}, implies that replacing
matter with antimatter preserves the relevant properties only when the
direction of time is also reversed. Within the present framework, this would
imply antisystemic properties of antimatter. Here we use an extended
understanding of CP and CPT invariance. In the traditional definition, these
symmetries remain within the boundaries of unitary dynamics and are
predominantly associated with fundamental properties of the relevant
Hamiltonians; here the definitions are extended to non-unitary evolution and
are associated with the \emph{covariant properties} of the \emph{Redfield}
and \emph{Lindblad} equations. We refer to the former as \emph{fundamental}
and to the latter as \emph{apparent} invariance.

The invariance properties of the Lindblad equation can, in general, be
affected both by the Hamiltonian $H$ and by the dephasing (or dissipative)
terms scaled by the coefficients $\gamma _{j}$ \cite%
{K-PhysA,SN-AS2021,Klimenko2026CDLindblad}. Since only small and rare
violations of CP invariance, associated with the Hamiltonians of weak
interactions, are known \cite{PDGConservation2020,PDG2024_RPP}, we shall
generally assume that the Hamiltonian is not only CPT-invariant but also
CP-invariant. Under CPT invariance, this also implies T-invariance of the
Hamiltonian. The situation is different for the dephasing terms, which are
not T-invariant. Therefore, the dephasing part of the Lindblad equation may
be either CP-invariant, corresponding to case I, or CPT-invariant,
corresponding to case II. We summarise this point in the proposition

\begin{proposition}
\label{Prop2}From the perspective of stochastic realism and intrinsic
dephasing, antimatter-sector dynamics may exhibit either systemic properties
associated with CP invariance or antisystemic properties associated with CPT
invariance. Since intrinsic dephasing terms are T-violating, the same
dephasing contribution cannot be simultaneously CP-invariant and
CPT-invariant.
\end{proposition}

Experimental selection between possibilities I and II may appear
straightforward: one would need to create a macroscopic, or at least
mesoscopic, amount of antimatter, isolate it completely from the systemic
influence of the environment, and then test its properties through
interactions with light. In practice, however, the situation is far more
complicated. Creating and retaining a substantial amount of antimatter, most
realistically antihydrogen, is extremely challenging \cite%
{Etesi2022,Klimenko2026Antisystems}. Moreover, if such antimatter were
produced and found to display antisystemic properties, it would be
thermodynamically unstable in a systemic environment, leading to substantial
disturbances and making controlled experimental observations rather
difficult. We need a more feasible alternative such as diffractive
dissociation.

\section{Apparent CP bias in proton--antiproton diffractive dissociations}

This section considers experiments examining \emph{diffractive dissociation} 
\cite{HEPDiffraction2002,DeWolf2002,PDGSoftQCD2021} in proton--proton ($pp$)
and proton--antiproton ($p\bar{p}$) collisions. These collisions involve
particles and antiparticles and, as discussed below, may be expected to
display some effective dissipative dynamics associated with the initial
stage of non-unitary evolution: dephasing, as the fragments formed in the
collisions lose coherence. In any case, these fragments do not form a single
thermalised strongly interacting medium, do not proceed to full
thermalisation, and therefore provide a useful setting for applying the
generalised dephasing Lindblad framework.

\begin{figure}[tbp]
\includegraphics[width=\textwidth,page=3,trim=1cm 1cm 1cm 1cm, clip
]{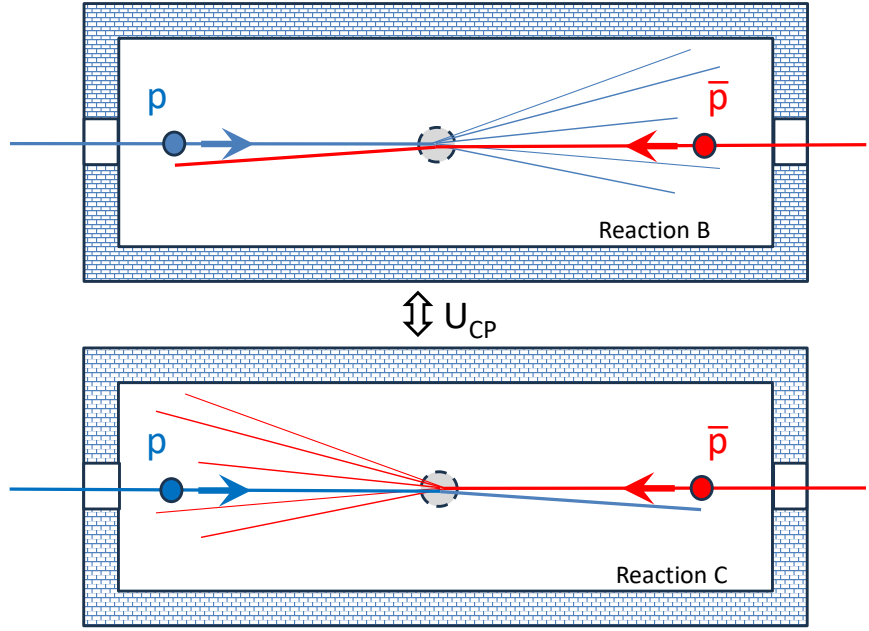}
\caption{Proton--antiproton collisions producing diffractive dissociation
reactions (B) and (C), shown as CP transformations of one another.}
\label{fig3}
\end{figure}

Our discussion here is focused on the following single-diffraction (SD)
dissociative reactions: 
\begin{gather}
\text{A)\ }p+p\rightarrow X+p  \label{rcc1} \\
\text{B)\ }p+\bar{p}\rightarrow X+\bar{p} \\
\text{C)\ }p+\bar{p}\rightarrow p+\bar{X}  \label{rcc2}
\end{gather}%
The notation is chosen in accordance with previous publications, where a
broader set of dissociative reactions is considered \cite%
{Klimenko2026CDLindblad}. The detailed compilation of diffractive data
indicates that 
\begin{equation}
\phi ^{2}=\frac{\sigma _{\mathrm{B}}}{\sigma _{\mathrm{A}}}=\frac{\sigma _{%
\mathrm{C}}}{\sigma _{\mathrm{B}}}  \label{fi2}
\end{equation}%
is $\phi ^{2}\approx 0.8$ when consistently evaluated from ISR, UA4, UA5,
CDF, D0, ALICE, and E710 experiments \cite%
{UA4Bernard1987,UA5Ansorge1986,CDF_Abe1994_SD,Pal2011_D0_FPD,ALICE_Abelev2013,E710_Amos1990,E710_Amos1993_SD}%
; essentially the same ratio $\phi ^{2}\approx 0.8$ is obtained when
comparing double diffraction in $p+\bar{p}$ and $p+p$ collisions. These
reactions are discussed in Ref. \cite{Klimenko2026CDLindblad} and are not
repeated here. Although $\phi ^{2}<1$ may seem conceptually surprising, the
value $\phi ^{2}\approx 0.8$ markedly improves the fit to the experimental
data with few fitting parameters. Such a value nonetheless calls for a
theoretical explanation. We group the possible explanations as follows,
regarding the first two as less likely.

\begin{description}
\item[Experimental errors] The value $\phi ^{2}\approx 0.8$ is determined
with a p-value $<5\times 10^{-4}$ when based on the reported experimental
uncertainties, and around $10^{-8}$ when based on the precision of the
approximation. Two conventional explanations must therefore be considered.
The first is that random errors conspire to form a consistent biased
pattern. This is possible but highly unlikely, precisely because the pattern
would have to be consistent. The second is a genuine systematic bias in the
measurements. Such a bias cannot be excluded in principle, but to explain
the effect it would have to reach $\sim 10\%$, act in the same direction,
and recur across multiple experiments performed by different groups, at
different locations and energies, among some of the most carefully executed
measurements in high-energy physics. Moreover, it would also have to
reproduce the consistent $\phi ^{n}$ pattern seen across single diffraction,
double diffraction, and the side-separated E710 counts. A systematic effect
of this kind remains possible, but establishing it would require identifying
a specific mechanism, correcting for it, and thereby showing that $\phi
^{2}<1$ no longer holds---effectively revising the outcomes of experimental
work conducted over several decades. Absent any identified cause or
mechanism, this possibility remains speculative.

\item[Fundamental violation] The immediate response from high-energy physics
would appeal to a subleading crossing-odd Reggeon contribution. While the
crossing-even Pomeron contributes equally to $pp$ and $p\bar{p}$ collisions
and is expected to dominate at high energies, subleading crossing-odd terms
can discriminate between them \cite%
{Pomeranchuk1958,DonnachieLandshoff1992,Collins1977_ReggeTheory,PDGSoftQCD2021,D0TOTEMOdderon2021}%
. This offers a reasonable explanation, grounded in the conventional unitary
dynamics of high-energy collisions, for why $\sigma _{\mathrm{B}}/\sigma _{%
\mathrm{A}}<1$. There is, however, a decisive problem: as illustrated in
Figure \ref{fig3}, reactions B and C are related by a CP transformation, so $%
\sigma _{\mathrm{C}}/\sigma _{\mathrm{B}}<1$ generally constitutes a direct
CP violation. Unlike the $pp$-versus-$p\bar{p}$ comparison, this ratio
cannot be affected by crossing-odd exchange, since B and C share the same
initial state. In this interpretation the effect would be a CP violation
associated with the fundamental properties of the particle Hamiltonians.
Such a violation is termed fundamental here and is regarded as highly
unlikely: CP violation in the strong interaction is known to be exceedingly
small, and an effect reaching $\sim 10\%$ in a strong-interaction process
would be inconsistent with everything established about the strong sector 
\cite{Symmetry1972,PDGConservation2020,PDG2024_RPP}.

\item[Apparent interferences] This explanation assumes fundamental CP
invariance of the underlying Hamiltonians and relates $\phi ^{2}<1$ to
interference from the environment or from intrinsic stochasticity. In this
context the observed bias is referred to as apparent and is attributed to
one of the following:

\begin{enumerate}
\item intrinsic dephasing terms, presumed to exist from the perspective of
stochastic realism; or

\item decoherence induced by environmental interference, associated with the
perspective of stochastic epistemicism.
\end{enumerate}
\end{description}

While experimental errors and fundamental CP violation are deemed unlikely,
our analysis points toward decoherence-related interference effects for two
reasons: the necessary presence of some form of effective decoherence or
channel selection in observed diffractive dissociation, and the extremely
high energy density involved in high-energy collisions.

According to the Good--Walker description of diffraction \cite%
{GoodWalker1960,DeWolf2002,Gustafson2012}, the incoming hadronic state $%
|p\rangle $ collides and is transformed by a transition operator $\hat{T}$,
which is considered to be Hermitian at Born order, although this Hermiticity
need not hold exactly at higher orders. The other states $|X_{k}\rangle $
are taken to be orthonormal mass eigenstates, while $|X_{0}\rangle
=|p\rangle $ is defined as the initial state. The transition amplitude
between $|p\rangle $ and the state $|X_{k}\rangle $ is given by $\hat{T}%
_{k0}=\langle X_{k}|\hat{T}|p\rangle $. Hence, at this order, the elastic
transition probability is 
\begin{equation}
P_{0}=\left\langle \mathrm{\hat{T}}\right\rangle _{p}^{2},\ \ \ \ \ \
\left\langle \mathrm{\hat{T}}\right\rangle _{p}\overset{\text{def}}{=}%
\left\langle p\right\vert \mathrm{\hat{T}}\left\vert p\right\rangle =\hat{T}%
_{00}
\end{equation}%
The application of the transition operator converts the initial state $%
|p\rangle $ into the transition state $|f\rangle $. In the basis $%
\{|X_{k}\rangle \}$, this state is given by the coherent superposition 
\begin{equation}
\left\vert f\right\rangle =\mathrm{\hat{T}}\left\vert p\right\rangle
=\sum_{k=0,1,...}\hat{T}_{k0}\left\vert X_{k}\right\rangle
\end{equation}%
In the traditional Good--Walker treatment, the coherent transition state is
projected onto observable, or partially observable, states $|X_{k}\rangle $
in accordance with the Born rule. Equivalently, one may assume that these
states decohere into the mixture \TEXTsymbol{\backslash}%
\begin{equation}
\rho _{t}=\sum_{k=0,1,...}P_{k}\left\vert X_{k}\right\rangle \left\langle
X_{k}\right\vert .
\end{equation}%
where $\rho _{t}$ denotes a generally unnormalised transition density. While
the first approach invokes measurement and is associated with stochastic
epistemicism, the second approach is closer to an ontic perspective. While
these assumptions are not identical, the outcome of both approaches is the
same for $P_{k}$: 
\begin{equation}
P_{k}=\left\vert \hat{T}_{k0}\right\vert ^{2}=\hat{T}_{k0}^{\ast }\hat{T}%
_{k0}
\end{equation}%
The overall probability of elastic plus diffractive transition, $%
P_{p\rightarrow pX}$, is then given by 
\begin{eqnarray}
P_{p\rightarrow pX} &=&\sum_{k=0,1,...}\hat{T}_{k0}^{\ast }\hat{T}_{k0} \\
&=&\sum_{k=0,1,...}\left\langle p\right\vert \hat{T}^{\dagger }\left\vert
X_{k}\right\rangle \left\langle X_{k}\right\vert \hat{T}\left\vert
p\right\rangle =\left\langle p\right\vert \hat{T}^{\dagger }\hat{T}|p\rangle
=\left\langle \mathrm{\hat{T}}^{2}\right\rangle _{p}  \notag
\end{eqnarray}%
where the final equality uses the Born-order Hermiticity of $\hat{T}$ and
assumes completeness $\Sigma _{k}\left\vert X_{k}\right\rangle \left\langle
X_{k}\right\vert =I$. This contains both the elastic probability $%
P_{p\rightarrow p}=P_{0}$ and the diffractive-scattering probability $%
P_{p\rightarrow X}$: 
\begin{equation}
P_{p\rightarrow X}=P_{p\rightarrow pX}-P_{p\rightarrow p}=\left\langle 
\mathrm{\hat{T}}^{2}\right\rangle _{p}-\left\langle \mathrm{\hat{T}}%
\right\rangle _{p}^{2}
\end{equation}%
Therefore, depending on philosophical perspective, this analysis can also be
interpreted as involving effective decoherence of the final physical states.
When a quantum formulation involves decoherence, the decoherence intensity
usually does have a direct effect on the transition rate and these effects
are to be taken into account in our analysis of dissociative reactions \cite%
{SN-AS2021,Klimenko2026CDLindblad}.

The other relevant factor is the extremely high energy density in
high-energy collisions, whereas the magnitude of the possible interferences
remains much smaller. It may seem unlikely that energetically weak processes
could affect processes involving extreme energy densities. There is,
however, an exception: disturbances related to the initiation of decoherence
and the arrow of time can be very small and still have an appreciable effect
on physical reality. For example, Penrose~\cite{Penrose2014a} assumed that
gravitational terms, which are extremely small for elementary particles, may
initiate non-unitary behaviour. Our analysis therefore points to the
possible relevance of effective non-unitary reduced dynamics to diffractive
dissociation:

\begin{proposition}
The apparent CP bias observed in diffractive dissociation is much more
plausibly associated with interference of non-unitary effects than with the
possibility of fundamental CP-violating terms in the unitary QCD Hamiltonian.
\end{proposition}

Since fundamental CP violation is unlikely under the specified conditions,
the observed asymmetry can be naturally interpreted, within the present
framework, as evidence for interference and non-unitary effects. This
motivates an analysis of the symmetry properties of the Lindblad equation
and of the corresponding dephasing models, with particular attention to
whether CP-asymmetric dephasing terms can arise while the associated
Hamiltonians remain CP-invariant. As noted above, the apparent CP bias or
violation may have two principal origins: first, intrinsic
decoherence/recoherence mechanisms constrained by CPT invariance, where the
time-asymmetry of stochastic dynamics can induce an effective CP asymmetry;
and second, an unbalanced environmental influence. These two mechanisms are
examined in the following sections. Environmental effects are considered
first because they involve multiple cases under various constraints, while
intrinsic effects are addressed in the following section.

\begin{figure}[tbp]
\includegraphics[width=\textwidth,page=4,trim=0cm 0cm 10cm 7cm, clip
]{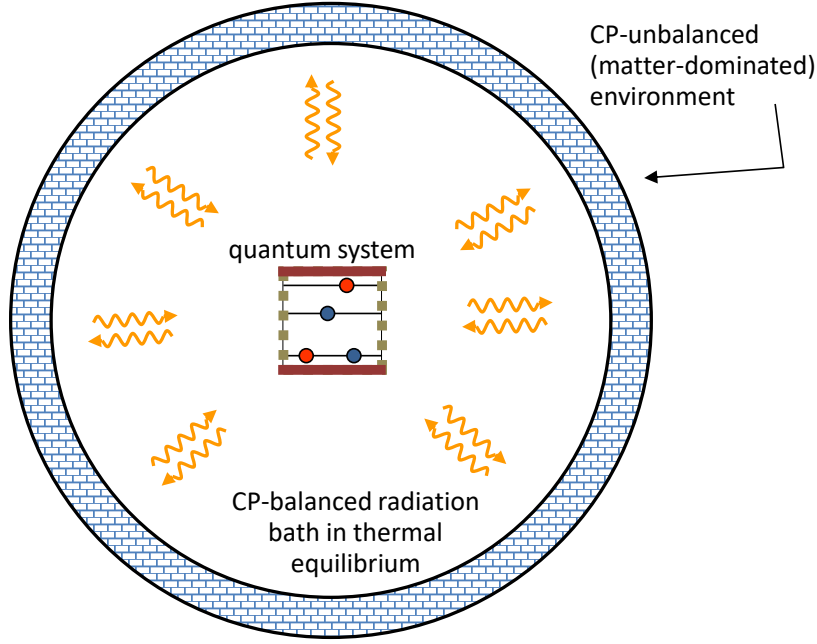}
\caption{A quantum system placed within matter-dominated surroundings and
interacting with its environment through a radiation bath. }
\label{fig4}
\end{figure}

\section{Can environmental interference create an apparent CP bias?}

A quantum system placed in an equilibrated bath can experience
time-directional interactions from the bath leading to decoherence in the
system \cite{Breuer2007book,LindbladIntro2020}. To analyse the covariance
properties of a quantum system placed in an environment, we consider a
three-tier interaction structure as suggested in Ref.~\cite{K-PhysA} and
depicted in Figure \ref{fig4}: the system $S$ embedded in an environment
consisting of a bath $B$ and the surroundings $G$. In the interaction chain $%
S\longleftrightarrow B\longleftrightarrow G$, the bath can serve as the
interaction agent mediating between the system and the wider surroundings.
Our consideration distinguishes \emph{fundamental} invariance, which is
characterised by the symmetry properties of the total Hamiltonian of the
system plus environment, from $\emph{apparent}$ invariance, which refers
only to the reduced dynamics observed at the level of the system $S$.

The joint Hamiltonian on the Hilbert space $\mathcal{H}_{S}\otimes \mathcal{H%
}_{B}\otimes \mathcal{H}_{G}$ is written as 
\begin{equation}
H=H_{S}\otimes I_{B}\otimes I_{G}+I_{S}\otimes H_{B}\otimes
I_{G}+I_{S}\otimes I_{B}\otimes H_{G}+h_{\text{int}}.  \label{eq:H_SBO}
\end{equation}%
where the $I$'s denote the corresponding identity operators. The interaction
Hamiltonian $h_{\text{int}}$ is decomposed into system--bath and
bath--surroundings interaction Hamiltonians $h_{SB}$ and $h_{BG}$ as 
\begin{equation}
h_{\text{int}}=h_{SB}\otimes I_{G}+I_{S}\otimes h_{BG},\ \
h_{SB}=\sum_{j}L_{j}\otimes V_{j}\   \label{eq:hSB-schmidt}
\end{equation}%
Note the absence of direct interactions between the system and the
surroundings, i.e. $h_{SG}=0$.

In the dephasing framework, all operators, ($H,$ $h,$ $L_{j},$ $V_{j},$ $%
\rho $), are presumed to be Hermitian, while all Hamiltonians ($H,$ $h$) are
CP- and CPT-invariant (i.e. no fundamental violations are present), for
example 
\begin{equation}
H=H^{\dagger },\ \ \ \mathrm{U}_{\text{CP}}H\mathrm{U}_{\text{CP}}^{\dagger
}=H,\ \ \Theta _{\text{CPT}}H{\Theta }_{\text{CPT}}^{-1}=H
\end{equation}%
where $\mathrm{U}_{\text{CP}}^{\dagger }=$ $\mathrm{U}_{\text{CP}}^{-1}$ due
to unitarity. \ The corresponding relation for ${\Theta }_{\text{CPT}}$
requires care since this operator is antiunitary. Since $\Theta _{\text{CPT}%
}=\mathrm{U}_{\text{CP}}\Theta _{\text{T}},$ the CP- and CPT-invariant
Hamiltonians are also T-invariant $\Theta _{\text{T}}H{\Theta }_{\text{T}%
}^{-1}=H.$

Note that the decomposition operators $L_{j}$ and $V_{j}$ are associated
with a CP-adapted operator Schmidt decomposition of the interaction
Hamiltonian in equation (\ref{eq:hSB-schmidt}) and a subsequent CP
decomposition discussed in Appendix \ref{AppC}. Here CP-adapted means that,
when the operator basis is not unique, it is chosen so that the relevant
factors have definite CP parity. This yields decomposition into CP-parity
operators which are not necessarily CP-even, that is 
\begin{equation}
\mathcal{U}_{\text{CP}}\left( L_{j}\otimes V_{j}\right) =+L_{j}\otimes V_{j}
\end{equation}%
but 
\begin{equation}
\mathcal{U}_{\text{CP}}(L_{j})=\zeta _{j}L_{j}\ \text{and }\ \mathcal{U}_{%
\text{CP}}(V_{j})=\zeta _{j}V_{j}
\end{equation}%
where $\zeta _{j}=\pm 1$ and $\mathcal{U}_{\text{CP}}(...)=\mathrm{U}_{\text{%
CP}}(...)\mathrm{U}_{\text{CP}}^{\dagger }$ is a linear superoperator
denoting CP transformation in the Hilbert space of the argument. Operator $%
L_{j}\otimes V_{j}$ is therefore \emph{CP-invariant} (i.e. CP-even) while
operators $L_{j}$ and $V_{j}$ are jointly either \emph{CP-even} $\zeta
_{j}=+1$ or \emph{CP-odd} $\zeta _{j}=-1$ (or, one can say, carry CP parity).

While $L_{j}=\sum_{n}\lambda _{jn}\left\vert n\right\rangle \left\langle
n\right\vert $ is always diagonal in pure dephasing (compare to (\ref{eq_H_L}%
)), the bath component of the interaction Hamiltonian can be diagonal 
\begin{equation}
V_{j}=\sum_{k}v_{jk}\left\vert k\right\rangle \left\langle k\right\vert 
\end{equation}%
(as, for example, in Zurek's dephasing model --see Appendix \ref{appBZ}) or
can be Hermitian non-diagonal in a general case. 

Consider a density matrix $\rho $ characterising some quantum system or
environment. It is called \emph{CP-balanced} if 
\begin{equation}
\mathcal{U}_{\text{CP}}\left( \rho \right) \overset{\text{def}}{=}\mathrm{U}%
_{\text{CP}}\rho \mathrm{U}_{\text{CP}}^{\dagger }\overset{\text{def}}{=}%
\bar{\rho}=\rho
\end{equation}%
i.e. if $\rho $ is CP-even. Notation $\bar{\rho}\overset{\text{def}}{=}%
\mathcal{U}_{\text{CP}}(\rho )$ is used here and throughout the paper. The
term is used to distinguish \emph{CP balance} of a state of a system from
its \emph{CP invariance} of the system Hamiltonian: while all Hamiltonians
are presumed to be CP-invariant, this may or may not be extended to the
density. Typically, the surrounding environment is matter-dominated and is
therefore \emph{CP-unbalanced} at the level of its density matrix, even when
the Hamiltonians governing that environment are CP-invariant.

From a fundamental perspective, CP conjugation of the system alone is
incomplete. A full CP transformation must also act on the environment,
converting matter into antimatter and vice versa in the bath, surroundings,
and the rest of the Universe \cite{K-PhysA}.

Joint quantum evolution of the system and environment is governed by von
Neumann equation%
\begin{equation}
\frac{d\rho (t)}{dt}=\mathcal{N}\left( \rho (t)\right) \overset{\text{def}}{=%
}-\frac{i}{\hbar }[H,\rho (t)]  \label{von-Neumann}
\end{equation}%
This equation is \emph{fundamentally CP-covariant} 
\begin{equation}
\frac{d\bar{\rho}(t)}{dt}=\mathcal{U}_{\text{CP}}^{\text{full}}\left( \frac{%
d\rho (t)}{dt}\right) =\mathcal{U}_{\text{CP}}^{\text{full}}\left( \mathcal{N%
}\left( \rho (t)\right) \right) =\mathcal{N}\left( \bar{\rho}(t)\right)
\label{von-N-cov}
\end{equation}%
when the overall Hamiltonian is CP-invariant $\mathcal{U}_{\text{CP}}^{\text{%
full}}(H)=H.$ The superscript is used to emphasise that the CP
transformation is applied to both the system and environment. This is
obvious in the present context but the scope of application needs clear
identification when applied to reduced description.

The Redfield/Lindblad-type generator $\mathcal{R}(\rho _{S})$ 
\begin{eqnarray}
\frac{d\rho _{S}(t)}{dt} &=&\mathcal{R}\left( \rho _{S};\rho _{BG}\right) =-%
\frac{i}{\hbar }[H_{S}+H_{\mathrm{eff}}(t),\rho _{S}(t)] \\
&&-\frac{1}{2\hbar ^{2}}\sum_{j,\ell }\gamma _{j\ell }\left[ L_{j},\left[
L_{\ell },\rho _{S}(t)\right] \right]  \notag
\end{eqnarray}%
which evolves the system density matrix $\rho _{S}(t)$ forward in time, is
considered in Appendix \ref{app:CPinSBG} and dependence of the model
coefficients $\gamma _{j\ell }=\gamma _{j\ell }(\rho _{BG})$ on $\rho _{BG}$%
, the density matrix of the environment, is shown explicitly in $\mathcal{R}%
\left( \rho _{S};\rho _{BG}\right) .$ The \emph{fundamental CP-invariance}
implies that 
\begin{equation}
\frac{d\bar{\rho}_{S}}{dt}=\mathcal{U}_{\text{CP}}^{\text{full}}\left( \frac{%
d\rho _{S}}{dt}\right) =\mathcal{U}_{\text{CP}}^{\text{full}}\left( \mathcal{%
R}\left( \rho _{S};\rho _{BG}\right) \right) =\mathcal{R}\left( \bar{\rho}%
_{S};\bar{\rho}_{BG}\right)  \label{Red-inv0}
\end{equation}%
---the simultaneous CP conjugation of the system and environment does not
change the evolution. Note that the Redfield equation is approximate, while
approximate equations may or may not be CP-covariant even if the overall
Hamiltonian is strictly preserved under CP transformation. As stated in
theorem \ref{theorA1}, the Redfield equation and, therefore, its direct
derivatives are all CP-covariant.\ 

The \emph{apparent CP invariance} is characterised by superoperator $%
\mathcal{U}_{\text{CP}}^{\text{app}}(...)$ that is applied only to the
system but not to the environment, that is symbolically 
\begin{equation}
\mathcal{U}_{\text{CP}}^{\text{full}}\left( \rho _{S}\otimes \rho
_{BG}\right) =\bar{\rho}_{S}\otimes \bar{\rho}_{BG}\text{ but }\mathcal{U}_{%
\text{CP}}^{\text{app}}\left( \rho _{S}\otimes \rho _{BG}\right) =\bar{\rho}%
_{S}\otimes \rho _{BG}  \label{fund-app}
\end{equation}%
--- the apparent CP transformation is strictly speaking incomplete. In terms
of the reduced description, the apparent CP invariance implies 
\begin{equation}
\frac{d\bar{\rho}_{S}}{dt}=\mathcal{U}_{\text{CP}}^{\text{app}}\left( \frac{%
d\rho _{S}}{dt}\right) =\mathcal{U}_{\text{CP}}^{\text{app}}\left( \mathcal{R%
}\left( \rho _{S};\rho _{BG}\right) \right) =\mathcal{R}\left( \bar{\rho}%
_{S};\rho _{BG}\right)  \label{Red-inv1}
\end{equation}%
which may or may not be correct. Indeed, Redfield-Lindblad operator involves
products $L_{j}L_{\ell }$ while their CP transformation yields 
\begin{equation}
\mathcal{U}_{\text{CP}}\left( L_{j}L_{\ell }\right) =\mathcal{U}_{\text{CP}%
}\left( L_{j}\right) \mathcal{U}_{\text{CP}}\left( L_{\ell }\right) =\zeta
_{j}\zeta _{\ell }L_{j}L_{\ell }
\end{equation}%
where $\zeta _{j}\zeta _{\ell }$ can be $-1$ when $L_{j}$ is CP-odd and $%
L_{\ell }$ is CP-even (or vice versa). The terms with $\zeta _{j}\zeta
_{\ell }=-1$ are apparently CP-violating. The sufficient conditions for
apparent CP invariance are given in Appendix \ref{AppC}.

Note that, unless a Redfield- or Lindblad-type generator reduces to purely
unitary evolution, it contains a non-unitary, time-directed environmental
contribution. Since this contribution is defined through antecedent causal
assumptions, which are time-directional, the autonomous dynamics of the
resulting reduced generator need not be CPT-invariant \cite%
{K-PhysA,Klimenko2016,Klimenko2026CDLindblad}.

A system or environment is called \emph{CP-neutral} when each of its states $%
\left\vert k\right\rangle $ is mapped onto itself by the CP transformation 
\begin{equation}
\mathrm{U}_{\text{CP}}\left\vert k\right\rangle =+\left\vert k\right\rangle
\end{equation}%
A system or environment which is not CP-neutral can nevertheless be
represented in a diagonally neutral form. A basis is called \emph{%
CP-diagonalised} when it is linked to energy eigenstates and 
\begin{equation}
\mathrm{U}_{\text{CP}}\left\vert k\right\rangle =\xi _{k}\left\vert
k\right\rangle ,\ \ \ \text{where }\xi _{k}=\pm 1
\end{equation}%
The details are given in Appendix \ref{AppC}.

In practice, the balance condition can be satisfied by the bath $\bar{\rho}%
_{B}=\rho _{B}$ but not necessarily by the whole environment, since,
typically, $\bar{\rho}_{G}\neq \rho _{G}$, as in predominantly
matter-populated surroundings. This naturally raises the question of whether
the influence of CP imbalance can propagate from the surroundings to the
system through the bath, or whether it is screened by the bath. The answer
to this question is not trivial and is given in Appendix \ref{AppC} and by
the theorem below.\ 

The bath may either transmit the CP imbalance of the surroundings to the
system---this case is referred to as \emph{CP-conductive}---or screen the
system from that influence---this case is referred to as \emph{CP-screening}%
---implying that the CP-unbalanced influence of the surroundings does not
affect the system, whose evolution remains CP-invariant (\ref{Red-inv1}).

\begin{theorem}
An equilibrated radiation bath cannot on its own induce apparent CP bias in
the system but it can be CP-conductive, passing CP bias from unbalanced
surroundings to the system. If, however, the bath-coupling operators $V_{j}$
are all diagonal in the CP-diagonalised basis, the CP imbalance of the
environment becomes entirely invisible to the system, which remains
apparently CP-invariant.
\end{theorem}

First, note that $V_{j}$ that are diagonal in the CP-diagonalised basis
cannot be CP-odd and, therefore, CP-odd operators $L_{j-}$ are absent as
demonstrated in lemma \ref{LemC4}.

Second, a radiation bath is not CP-neutral (and therefore is not restricted
by lemma \ref{lemC3}), at least because magnetic field changes its sign
under CP transformation: charge transformation reverses its direction but
parity preserves it since magnetic field is axial.

Third, an equilibrium radiation bath with a CP-invariant Hamiltonian is
CP-balanced. Indeed, the density matrix of radiation equilibrated at
temperature $T$ is determined by the Gibbs state,%
\begin{equation}
\rho _{B}^{\text{eq}}=\frac{\exp \left( -\frac{H_{B}}{k_{\text{B}}T}\right) 
}{Z_{T}},\ \ \ Z_{T}=\func{Tr}\left( \exp \left( -\frac{H_{B}}{k_{\text{B}}T}%
\right) \right)
\end{equation}%
where $k_{\text{B}}$ is the Boltzmann constant. Since $H_{B}$ is
CP-invariant (and therefore CP-even) $\mathcal{U}_{\text{CP}}(H_{B})=+H_{B},$
the same relation applies to its exponential. Therefore, 
\begin{equation}
\mathcal{U}_{\text{CP}}\left( \rho _{B}^{\text{eq}}\right) =+\rho _{B}^{%
\text{eq}}
\end{equation}%
and the equilibrated bath is CP-balanced. It is presumed that the bath
remains in its equilibrium state. Note that if surroundings are absent then $%
\rho _{BG}=\rho _{B}=\rho _{B}^{\text{eq}}=\bar{\rho}_{B}=\bar{\rho}_{BG}$
and the Redfield/Lindblad-type generator remains CP-invariant according to
lemmas \ref{lemC1} and \ref{lemC2}.

Among other possible examples, gravity is CP-neutral and, therefore, is
always CP-balanced. Higgs bosons are CP-neutral whereas gluons can produce
both CP-even exchanges (e.g. Pomerons) and CP-odd configurations, although
the latter correspond to subleading exchanges suppressed at high energies.
An equilibrium radiation bath does not induce CP asymmetry in the system on
its own but it can act as a CP-conductive medium and transmit such an
asymmetry if the asymmetry is already present in the surroundings. By
contrast, anisotropic electromagnetic environment, for example a fixed
external magnetic field, can induce apparent CP bias in the system.

In high-energy collisions, however, this environmental mechanism is
constrained by causality. Since the particle velocities are close to the
speed of light and the collision times are very short, it is not obvious how
distant surroundings could influence the interaction region during the
collision: a disturbance propagating at the speed of light from the
surroundings would generally require more time to reach the system. Thus, if
the system interacts only with an equilibrium radiation bath and not with
CP-asymmetric surroundings, the bath does not break CP invariance of the
system.

In principle, environmental effects can be tested experimentally by changing
the surroundings and checking whether the apparent CP bias in the system
changes systematically \cite{SN-AS2021}.

\section{Interplay of intrinsic and environmental dephasing}

Although remote environmental interactions capable of inducing apparent CP
bias are possible in high-energy diffractive dissociations, they are subject
to causality constraints and their presence and role require further
experimental clarification. The other possible source of apparent CP bias,
not associated with CP violations in true Hamiltonians, is \emph{intrinsic
randomness}, which is inherently associated with any quantum system
according to the perspective of \emph{stochastic realism} \cite%
{Klimenko2025TwoTypesTemporalSymmetry}. In this section we consider the
combined effect of both contributions, under the assumption that the
environmental influence preserves CP symmetry while the intrinsic
contribution may preserve either CP or CPT.

As stated in proposition~\ref{Prop2}, intrinsic randomness can be either CP-
or CPT-preserving. If both intrinsic and environmental dephasing preserve
CP, they share the same symmetry and no difference between matter and
antimatter is visible in experiments---the measurement of $\phi ^{2}$ does
not allow one to distinguish intrinsic decoherence from environmental
decoherence. If, however, intrinsic randomness is CPT-preserving while the
environmental contribution remains CP-preserving, the two mechanisms differ
in their action on matter versus antimatter, and the difference becomes
experimentally detectable.

Combining the intrinsic dephasing equation~(\ref{eq_Lind_nm}) and the
environmental dephasing equation~(\ref{eq_Red_nm}), the effective dephasing
rates are $\Gamma ^{B}+\Gamma ^{S}$ for matter and $\Gamma ^{B}\pm \Gamma
^{S}$ for antimatter, where the upper sign corresponds to CP-invariant
intrinsic dephasing and the lower sign to CPT-invariant intrinsic dephasing.
Here $\Gamma ^{B}$ is the dephasing rate associated with environmental
interference of the bath and $\Gamma ^{S}$ is the dephasing (or rephasing)
rate associated with intrinsic stochastic behaviour of the system; both are
taken positive. The view taken in Ref. \cite{Klimenko2026CDLindblad} is that
dissociation into the particle system $X$ involves a loss of coherence
between these particles, so that they cannot reassemble into the original
state $p$. For non-resonant processes, decoherence enhances the transition
rate; the same positive dependence of the transition probability on the
decoherence rate is assumed here. Assuming that the transition rate depends
on the dephasing rate and, in a linear approximation, is proportional to it,
the relative dephasing ratio $\phi ^{2}$ becomes 
\begin{equation}
\phi ^{2}=\frac{\sigma _{\mathrm{C}}}{\sigma _{\mathrm{B}}}=%
\begin{cases}
\dfrac{\Gamma ^{B}+\Gamma ^{S}}{\Gamma ^{B}+\Gamma ^{S}}=1, & \text{%
CP-invariant case}, \\[6pt] 
\dfrac{\Gamma ^{B}-\Gamma ^{S}}{\Gamma ^{B}+\Gamma ^{S}}<1, & \text{%
CPT-invariant case},%
\end{cases}
\label{eq:phi-sd-ratio}
\end{equation}%
since the signs of $\Gamma ^{B}$ and $\Gamma ^{S}$ coincide under CP
symmetry and are opposite under CPT symmetry. Note that the overall
dephasing rate $\Gamma ^{B}-\Gamma ^{S}$ remains positive, i.e. dominant
rephasing is not observed in these experiments.

Under \emph{CPT symmetry} of the intrinsic stochastic interferences, the
situation can be understood physically as follows. If antimatter has
antisystemic properties, then the intrinsic stochastic mechanism acting
inside the antiproton is not decoherence-inducing, but recoherence-inducing.
It therefore counteracts the dissociation in the system, because
dissociation requires loss of coherence between the outgoing hadronised
components. At the same time, both reactions are exposed to ordinary
environmental interaction-induced disturbances, which act in the usual
decoherence-producing direction. Thus, proton dissociation in reaction~B is
supported both by intrinsic systemic decoherence and by environmental
decoherence, whereas antiproton dissociation in reaction~C is affected by
two competing tendencies: \emph{environmental decoherence} promotes
dissociation, while \emph{intrinsic decoherence} suppresses it. The observed
value of $\phi ^{2}$ then reflects the competition between these two effects.

Within the present phenomenological interpretation, the deviation 
\begin{equation}
1 - \phi^{2} = \frac{2\Gamma^{S}}{\Gamma^{B} + \Gamma^{S}}
\label{eq:phi-deviation}
\end{equation}
measures the intrinsic contribution relative to the total dephasing rate and
may be used as an indicator of the relative significance of \emph{intrinsic
CPT-invariant dephasing mechanisms}.

We note that the diffractive dissociations do not display \emph{antisystemic
behaviour}---under experimental conditions the decohering effects remain
dominant. We observe only a significant but still modest $\Gamma ^{S}/\Gamma
^{B}\sim 10\%$ relative reduction in decohering intensity, indicating that
under the conditions of the experiment the environmental effects still
prevail \cite{Klimenko2026CDLindblad}. This situation may change as the
number of antiprotons increases further \cite{Klimenko2026Antisystems}, for
example in $\bar{p}\bar{p}$ collisions, although such experiments have not
yet been conducted. If recoherence were to become dominant, a qualitative
change in thermodynamic properties would be expected. The projected effect,
however, is that systemic behaviour would weaken further under such
conditions but would still prevail overall. Even if physically possible,
observing fully antisystemic behaviour may require even higher quantity and
density of antimatter than has been experimentally accessible to date.

\section{Discussion}

Figure~\ref{fig3} illustrates a diffractive $p\bar{p}$ collision with
reaction B at the top and reaction C at the bottom. A CP transformation
converts B into C, and vice versa; therefore, exact CP symmetry would not
allow these reactions to have different cross-sections. Experimentally,
however, a difference $1-\phi ^{2}\approx 20\%$ is observed. As discussed
previously \cite{Klimenko2026CDLindblad}, this discrepancy is not readily
attributable to experimental errors. At the same time, a fundamental CP
violation of this magnitude is equally unlikely. One is therefore led to
consider whether dissociative diffraction involves additional interference
or decoherence mechanisms.

Such interference can be induced by non-equilibrium radiation or a
directional electromagnetic field. For example, an external magnetic field $%
\mathbf{\vec{B}}_{\text{ext}}$ might be present in $p\bar{p}$ collisions.
Under CP transformation the magnetic field changes sign, $\mathcal{U}_{\text{%
CP}}(\mathbf{\vec{B}}_{\text{ext}}\mathbf{)=-\vec{B}}_{\text{ext}},$ so an
untransformed external field could make the two apparently CP-conjugate
experimental situations physically different. However, such a field cannot
plausibly explain the effect through its energy density alone. The energy
density of ordinary external magnetic fields is enormously smaller than the
energy density involved in high-energy hadronic collisions. It is therefore
unrealistic to assume that a field weaker by many orders of magnitude can
directly affect the collision probability at the $10\%$ level.

If external fields play any role, they must do so as small symmetry-breaking
seeds within a highly sensitive decoherence process. In dissociative
diffraction, the outgoing fragments must lose phase coherence; otherwise the
state would recombine coherently into the original particle channel. Such
decoherence may involve large amplification factors, making the process
sensitive to very small environmental or intrinsic perturbations.

Although models of gravitationally induced decoherence have been proposed 
\cite{Penrose2014a}, ordinary gravitational coupling is CP-neutral and
therefore cannot transmit a CP imbalance from the surroundings to the
system. In this context, radiation is a more plausible carrier of
environmental bias, because electromagnetic fields transform nontrivially
under CP. Radiation itself does not originate a CP bias --- equilibrated
radiation is omnipresent yet carries no matter--antimatter imbalance --- but
it can transmit a bias that originates from such an imbalance in the
surroundings. To induce an apparent CP bias, the radiative environment would
therefore have to couple to both the surroundings and the system.

Given the very short collision times, it is not clear how interactions
between surrounding matter and the system, mediated through radiation, could
influence the collision within causality constraints. This raises a \emph{%
prima facie} difficulty rather than a decisive objection, and the question
needs to be settled experimentally. The possibility of environmental
influence therefore cannot be discarded and requires further demonstration:
for example, if changing the arrangement of the surroundings changes the
magnitude of the observed bias, this would provide direct evidence for an
environmental origin of the bias.

It is also possible that each moving particle is surrounded by a
non-equilibrium electromagnetic \textquotedblleft
dressing\textquotedblright\ that remains dormant under unitary interactions
but reveals its presence whenever non-unitary effects become significant. At
the time of collision, these factors are already present and are enacted
immediately. Such possibilities should not be excluded, but at this stage
they remain speculative. Although \textquotedblleft
dressing\textquotedblright\ is treated here as an environmental effect, it
is effectively close to intrinsic explanation.

At the same time, the observed consistency of the bias across many
experiments and conditions points to a fixed intrinsic characteristic and,
if this is accepted as the explanation, to its CPT invariance. Under the
specific conditions considered here, the intrinsic contribution is
significant but the environmental influence still accounts for the dominant
share (i.e. $\Gamma ^{B}/(\Gamma ^{B}+\Gamma ^{S})\approx 90\%$
corresponding to $\phi ^{2}\approx 0.8$) of the effect, while the
possibility of forming true antisystems under intrinsic prevalence remains
uncertain. Further theoretical and experimental work is therefore needed,
and such work may have to go beyond the limitations of conventional unitary
dynamics.

\section{Conclusion}

High-energy diffractive proton--antiproton collision experiments, among the
most carefully executed measurements in physics, display a statistically
significant CP bias: about $10\%$ when expressed as a dephasing-rate
anomaly, or about $20\%$ when expressed as a cross-section ratio. This bias
cannot simply be dismissed as experimental inaccuracy. An unidentified
systematic effect common to different measurements, detector set-ups,
laboratories, and data-processing procedures remains possible in principle,
but it would have to act coherently across independent experiments and
reproduce the same decoherence factor $\phi $ in both single- and
double-diffractive observables. In the absence of a concrete mechanism of
this kind, such an explanation appears unlikely; the bias therefore calls
for a plausible and experimentally testable account.

The cross-section asymmetry detected in diffractive dissociation is
interpreted here as an apparent CP bias rather than a fundamental CP
violation: it is not attributed to CP violation of the true unitary
Hamiltonians, but instead to CP-asymmetric non-unitary interferences, which
must therefore be taken into account. This apparent bias can be attributed
to two distinct sources, depending on one's interpretive framework:
environmental interference, favoured by the perspective of stochastic
epistemicism, or intrinsic randomness, favoured by the perspective of
stochastic realism. These two perspectives are not mutually exclusive and
may both contribute to the observed effect.

If the environmental interference is mediated by an equilibrated radiation
bath, such an environment can allow an external CP bias to propagate through
the bath into the system, initiating an apparent CP bias of environmental
origin. Despite causal constraints, these mechanisms of environmentally
induced apparent CP bias should not be dismissed a priori and warrant active
experimental investigation. In particular, non-equilibrium or anisotropic
environments could generate CP-asymmetric dephasing.

Within the intrinsic branch of the present framework, the existing evidence
is consistent with CPT invariance of the intrinsic stochasticity, suggesting
that an intrinsic non-unitary contribution tends to respect CPT symmetry
even while appearing CP-asymmetric at the level of the reduced dynamics.
Thus the observed asymmetry is consistent with the broader non-unitary
dephasing/recoherence framework developed here, while leaving open whether
the relevant CP-asymmetric effective dynamics is environmental, intrinsic,
or a combination of both.

It is clear that the observed bias in proton--antiproton collisions is
puzzling and requires an explanation. The author's current view, based on
the available experimental data, is that the bias reflects intrinsic
stochastic behaviour priming the arrow of time. Other explanations remain
possible, however, and may become clearer as further theoretical and
experimental work develops.

\appendix

\section{Dual parabolicity and Conditional Moment Closure \label{app:CMC}}

One can distinguish three different meanings of dual parabolicity,
understood here as the simultaneous presence of both forward-parabolic and
backward-parabolic terms:

\begin{itemize}
\item \textbf{First}, in the odd-symmetric case, the forward and reverse
formulations describe the same stochastic process, and the chosen time
direction is irrelevant: the two descriptions are equivalent representations
of one time-direction-invariant law. The two formulations are associated
with opposite entropy behaviours.

\item \textbf{Second}, the same stochastic process may admit forward and
reverse formulations that are not equivalent; in this case the time
direction matters, and the directions of time are inherently inequivalent.

\item \textbf{Third}, a dual-parabolic structure may arise for reasons
unrelated to temporal reversal, when two different physical mechanisms, or
two different statistical aspects of one physical mechanism, are coupled in
the same formulation. Although such cases are less common, there is a
well-known example---Conditional Moment Closure (CMC)---which is explained
below.
\end{itemize}

An example of an equation with dual parabolicity that is not based on
temporal reversal, and therefore is not physically odd-symmetric, is given
by the divergence form of CMC \cite{Klimenko99}, which, in a simplified
case, is governed by the following equation: 
\begin{equation}
\frac{\partial (Q_{z}P_{z})}{\partial t}+\nabla \cdot (\mathbf{u}%
_{z}Q_{z}P_{z})=\frac{\partial }{\partial z}\left( N_{z}P_{z}\frac{\partial
Q_{z}}{\partial z}-Q_{z}\frac{\partial (N_{z}P_{z})}{\partial z}\right)
\label{CMC-div}
\end{equation}%
Here $Q_{z}$ is the conditional expectation of a reacting scalar $Y$ at a
fixed value of the mixture fraction $Z$, that is, $Q_{z}=\langle
Y|Z=z\rangle $. The equation also involves the probability density function
(PDF) of the mixture fraction, $P_{z}$, the conditional velocity, $\mathbf{u}%
_{z}$, and the conditional scalar dissipation, $N_{z}$. Note that, despite
the strong formal similarity $Q_{z}\sim \varphi $ and $P_{z}\sim \psi $ with
equation (\ref{FP-f}), which governs the laws of nature under the
stochastic-realist interpretation with dual temporal boundary conditions, $%
Q_{z}$ and $P_{z}$ are only adjoint quantities. Unlike $\varphi $ and $\psi $
under odd-symmetric conditions, they are not physically equivalent.

The two parabolic contributions represent two different processes associated
with turbulent mixing in high-Reynolds-number flows. They were derived
independently, based on different physical assumptions and without invoking
time reversal \cite{Klimenko99}. The relation of these equations to time
reversal was identified later \cite{K_QJMAM}. CMC provides a physically
meaningful example in which dual parabolicity reflects coupled
physical/statistical mechanisms rather than temporal reversal. Even if $%
\func{div}(\mathbf{u}_{z})=0$ and $N_{z}$ does not depend on $z$, these
equations are not physically time-symmetric (odd-symmetric), because both
processes---turbulent diffusion of $Z$ and turbulent diffusion of $Y$%
---increase physical entropy in forward time $t$.

\section{Reduced description of system--bath--surroundings interactions 
\label{app:CPinSBG}}

\subsection{Second-order approximation}

The overall Hamiltonian $H$ can be represented equivalently as describing
interactions between a system, represented by the Hilbert space $\mathcal{H}%
_{S}$, and an environment consisting of the bath and the surroundings,
represented by the Hilbert space $\mathcal{H}_{B}\otimes \mathcal{H}_{G}$: 
\begin{eqnarray}
H &=&H_{0}+h_{I},\ \ \ \ H_{0}=H_{S}\otimes I_{BG}+I_{S}\otimes H_{BG}
\label{eq:Htotal} \\
h_{I} &=&h_{SB}\otimes I_{G}=\sum_{j}L_{j}\otimes V_{j}\otimes
I_{G}=\sum_{j}L_{j}\otimes W_{j}  \label{eq:hSB}
\end{eqnarray}%
where%
\begin{eqnarray}
H_{BG} &=&H_{B}\otimes I_{G}+I_{B}\otimes H_{G}+h_{BG}=H_{BG}^{(0)}+h_{BG} \\
h_{BG} &=&\sum_{\beta }B_{\beta }\otimes G_{\beta }\   \label{eq:hBG}
\end{eqnarray}%
is the environmental (bath and surroundings) Hamiltonian.

The perturbation series up to second order of the von Neumann equation (\ref%
{von-Neumann}) can be written in the Schr\"{o}dinger picture as \cite%
{Breuer2007book} 
\begin{eqnarray}
\frac{d\rho (t)}{dt} &=&\mathcal{N}\left( \rho (t)\right) =\mathcal{N}%
^{(0)}\left( \rho (t)\right) +\mathcal{N}^{(1)}\left( \rho (t)\right) +%
\mathcal{N}^{(2)}\left( \rho (t)\right) +... \\
\mathcal{N}^{(0)}\left( \rho (t)\right) &=&-\frac{i}{\hbar }[H_{0},\rho (t)]
\notag \\
\mathcal{N}^{(1)}\left( \rho (t)\right) &=&-\frac{i}{\hbar }\left[
h_{I},\rho ^{(0)}(t)\right]  \notag \\
\mathcal{N}^{(2)}\left( \rho (t)\right) &=&-\frac{1}{\hbar ^{2}}%
\int_{0}^{t}ds\left[ h_{I},U_{0}(t-s)[h_{I},\rho (s)]U_{0}^{\dagger }(t-s)%
\right] .  \notag
\end{eqnarray}%
Here $\rho ^{(0)}$ denotes 
\begin{equation}
\rho ^{(0)}(t)=U_{0}(t)\rho (0)U_{0}^{\dagger }(t),\ \ \ U_{0}(t)=\exp
(-iH_{0}t/\hbar )
\end{equation}%
the leading-order evolution of the density operator in the overall space.

\subsection{Born approximation and Redfield equation}

For pure dephasing, we assume 
\begin{equation}
\lbrack H_{S},L_{j}]=0,
\end{equation}%
so that 
\begin{equation}
L_{j}^{(0)}(-\tau )\overset{\text{def}}{=}e^{-iH_{S}\tau /\hbar
}L_{j}e^{iH_{S}\tau /\hbar }=L_{j}.
\end{equation}%
We also assume that all coupling operators are chosen Hermitian, which can
always be done when representing Hermitian $h_{SB}$ and $h_{BG}$: 
\begin{equation}
L_{j}^{\dagger }=L_{j},\qquad V_{j}^{\dagger }=V_{j},\qquad W_{j}^{\dagger
}=W_{j}.
\end{equation}

With these transformations, the \emph{Redfield equation} \cite%
{Redfield1957Relaxation,Breuer2007book} for the system density $\rho _{S}$,
obtained by tracing out the environment, $\rho _{S}=\func{Tr}_{BG}(\rho )$,
under the Born approximation 
\begin{equation}
\rho \approx \rho _{S}\otimes \rho _{BG}
\end{equation}%
and some higher order adjustments between the first and the second order
terms, becomes 
\begin{equation}
\frac{d\rho _{S}(t)}{dt}=\mathcal{R}\left( \rho _{S}(t)\right) =-\frac{i}{%
\hbar }[H_{S},\rho _{S}(t)]+\mathcal{R}^{(1)}\left( \rho _{S}(t)\right) +%
\mathcal{R}^{(2)}\left( \rho _{S}(t)\right)  \label{red1}
\end{equation}%
\begin{eqnarray*}
\mathcal{R}^{(1)}\left( \rho _{S}(t);\rho _{BG}\right) &=&-\frac{i}{\hbar }%
\sum_{j}\mu _{j}[L_{j},\rho _{S}(t)] \\
\mathcal{R}^{(2)}\left( \rho _{S}(t);\rho _{BG}\right) &=&-\frac{1}{\hbar
^{2}}\sum_{j,\ell }\int_{0}^{t}d\tau \left( C_{j\ell }^{\prime }(\tau
)[L_{j},L_{\ell }\rho _{S}(t)]+C_{j\ell }^{\prime \ast }(\tau )[\rho
_{S}(t)L_{\ell },L_{j}]\right) \\
&=&-\frac{1}{\hbar ^{2}}\sum_{j,\ell }\left( A_{j\ell }^{\prime
}(t)[L_{j},L_{\ell }\rho _{S}(t)]+A_{j\ell }^{\prime \ast }(t)[\rho
_{S}(t)L_{\ell },L_{j}]\right)
\end{eqnarray*}%
where%
\begin{equation}
A_{j\ell }^{\prime }(t)=\int_{0}^{t}C_{j\ell }^{\prime }(\tau )d\tau
\label{rede1}
\end{equation}%
is the time-integrated fluctuation correlation and $A_{j\ell }^{\prime \ast
}(t)$ is its complex conjugate. Here and in the rest of the paper, the
second argument $\rho _{BG}$ is added $\mathcal{R}\left( \rho _{S}\right) =%
\mathcal{R}\left( \rho _{S};\rho _{BG}\right) $ to emphasise the dependence
of the Redfield operator and its coefficients on $\rho _{BG}$. The
superscript "$\ast $" denotes complex conjugation, while the coefficients in
Redfield equation are given by%
\begin{gather}
C_{j\ell }(\tau )\overset{\text{def}}{=}\func{Tr}_{BG}\left( W_{j}W_{\ell
}^{(0)}(-\tau )\rho _{BG}\right) =\func{Tr}_{BG}\left( W_{\ell }^{(0)}(-\tau
)W_{j}\rho _{BG}\right) ^{\ast } \\
W_{j}^{(0)}(-\tau )\overset{\text{def}}{=}e^{-iH_{BG}\tau /\hbar
}W_{j}e^{iH_{BG}\tau /\hbar } \\
\mu _{j}\overset{\text{def}}{=}\func{Tr}_{BG}(W_{j}\rho _{BG})=\func{Tr}%
_{B}(V_{j}\rho _{B}) \\
C_{j\ell }^{\prime }(\tau )\overset{\text{def}}{=}C_{j\ell }(\tau )-\mu
_{j}\mu _{\ell },\ \ \ \ \rho _{B}\overset{\text{def}}{=}\func{Tr}_{G}\left(
\rho _{BG}\right) \ 
\end{gather}%
Assuming that the environment is stationary 
\begin{equation}
\lbrack H_{BG},\rho _{BG}]=0  \label{rede2}
\end{equation}%
leads to equation $C_{\ell j}^{\prime }(-\tau )=C_{j\ell }^{\prime \ast
}(\tau )$ that is used to relate terms in $\mathcal{R}^{(2)}$ to each other.

Assuming that both interaction Hamiltonians, $h_{SB}$ and $h_{BG}$, are weak
and of similar perturbative order, the drift and dissipation induced by the
bath are $\sim h_{SB}$ and $\sim h_{SB}^{2},$ respectively, i.e. of first
and second order. Interference between the system and the surroundings
through environmental drift is of order $\sim h_{SB}h_{BG}$. The leading
dissipative interference from the surroundings is generally of order $\sim
h_{SB}^{2}h_{BG}$; if the first-order drift correction due to interactions
between the surroundings and the bath vanishes, the leading contribution is
then $\sim h_{SB}^{2}h_{BG}^{2}$.

\subsection{Lindblad-GKS dephasing equation}

With the definitions 
\begin{equation}
\gamma _{j\ell }(t)\overset{\text{def}}{=}A_{j\ell }^{\prime }(t)+A_{\ell
j}^{\prime \ast }(t),\ \ \ \eta _{j\ell }(t)\overset{\text{def}}{=}\frac{%
A_{j\ell }^{\prime }(t)-A_{\ell j}^{\prime \ast }(t)}{2i}
\end{equation}%
the Redfield equation \cite{Redfield1957Relaxation,Breuer2007book} then
becomes 
\begin{equation}
\frac{d\rho _{S}(t)}{dt}=\mathcal{R}\left( \rho _{S}\right) =-\frac{i}{\hbar 
}[H_{S}+H_{\mathrm{eff}}(t),\rho _{S}(t)]+\mathcal{R}^{\prime }\left( \rho
_{S}\right)  \label{Redfield}
\end{equation}%
with Lindblad dissipator written in the GKS form \cite%
{Lindblad1976,GoriniKossakowskiSudarshan1976,Breuer2007book,LindbladIntro2020}
\begin{equation}
\mathcal{R}^{\prime }\left( \rho _{S}\right) \overset{\text{def}}{=}-\frac{1%
}{2\hbar ^{2}}\sum_{j,\ell }\gamma _{j\ell }(t)\left( \left\{ L_{j}L_{\ell
},\rho _{S}\right\} -2L_{\ell }\rho _{S}L_{j}\right)
\end{equation}%
for Hermitian $L_{j}^{\dagger }=L_{j}$. Here, the effective drift
Hamiltonian is given by 
\begin{equation}
H_{\mathrm{eff}}(t)=\sum_{j}\mu _{j}L_{j}+\frac{1}{\hbar }\sum_{j,\ell }\eta
_{j\ell }(t)L_{j}L_{\ell }  \label{eq:Heff-drift}
\end{equation}%
where the first term is of the first order and the second term---the Lamb
drift---is of the second order.

When $H_{BG}$, $W_{j}$ and $\rho _{BG}$ are real (and, therefore,
T-invariant in the spinless case) and $\rho _{BG}$ is in a steady state, the
matrices $C_{j\ell }=C_{\ell j},$ $C_{j\ell }^{\prime }=C_{\ell j}^{\prime }$%
, $A_{j\ell }=A_{\ell j}$ and $A_{j\ell }^{\prime }=A_{\ell j}^{\prime }$
become symmetric, so that $\gamma _{j\ell }$ and $\eta _{j\ell }$ are real;
since they must also be Hermitian, they are symmetric: $\gamma _{j\ell
}=\gamma _{\ell j}$ and $\eta _{j\ell }=\eta _{\ell j}$. Indeed, 
\begin{eqnarray}
C_{j\ell }(t) &=&\func{Tr}_{BG}\left( \left( W_{j}W_{\ell }^{(0)}(-t)\rho
_{BG}\right) ^{\text{{\tiny T}}}\right) =\func{Tr}_{BG}\left( \rho
_{BG}W_{\ell }^{(0)}(t)W_{j}\right)  \notag \\
&=&\func{Tr}_{BG}\left( W_{\ell }^{(0)}(t)W_{j}\rho _{BG}\right) =\func{Tr}%
_{BG}\left( W_{\ell }W_{j}^{(0)}(-t)\rho _{BG}\right) =C_{\ell j}(t)
\end{eqnarray}%
since $\rho _{BG}^{\text{{\tiny T}}}=\rho _{BG},$ $W_{j}^{\text{{\tiny T}}%
}=W_{j}$ and $(W_{\ell }^{(0)}(-t))^{\text{{\tiny T}}}=W_{\ell }^{(0)}(t)$,
where the superscript "{\small T}" denotes transposition. Under these
conditions, the dephasing term takes a double-commutator form 
\begin{eqnarray}
\mathcal{R}^{\prime }\left( \rho _{S}\right) &=&-\frac{1}{2\hbar ^{2}}%
\sum_{j,\ell }\gamma _{j\ell }(t)\left[ L_{j},\left[ L_{\ell },\rho _{S}(t)%
\right] \right] \\
&=&-\frac{1}{2\hbar ^{2}}\sum_{j}\gamma _{j}^{\circ }(t)\left[ L_{j}^{\circ
},\left[ L_{j}^{\circ },\rho _{S}(t)\right] \right]  \notag
\end{eqnarray}%
and can be diagonalised as shown, similarly to equation (\ref{eq:rho}).

Assume that the energy basis and the decoherence basis coincide: 
\begin{equation}
H_{S}|n\rangle =E_{n}|n\rangle ,\qquad L_{j}|n\rangle =\lambda
_{jn}|n\rangle .
\end{equation}

In the energy (and dephasing) basis $\rho _{nm}=\left\langle n\right\vert
\rho _{S}\left\vert m\right\rangle $, the Redfield equation becomes 
\begin{equation}
\frac{d\rho _{nm}(t)}{dt}=-\left( \frac{i}{\hbar }\Delta E_{nm}+\frac{i}{%
\hbar }M_{nm}+\frac{i}{\hbar ^{2}}\Lambda _{nm}(t)+\frac{\Gamma _{nm}^{B}(t)%
}{\hbar ^{2}}\right) \rho _{nm}(t).  \label{eq_Red_nm}
\end{equation}%
where%
\begin{equation}
\Delta E_{nm}=E_{n}-E_{m},\ \ \ \Gamma _{nm}^{B}(t)=\frac{1}{2}\sum_{j,\ell
}\gamma _{j\ell }(t)(\lambda _{jn}-\lambda _{jm})(\lambda _{\ell n}-\lambda
_{\ell m})
\end{equation}%
\begin{equation}
M_{nm}=\sum_{j}\mu _{j}(\lambda _{jn}-\lambda _{jm}),\ \ \ \Lambda
_{nm}(t)=\sum_{j,\ell }\eta _{j\ell }(t)\left( \lambda _{jn}\lambda _{\ell
n}-\lambda _{jm}\lambda _{\ell m}\right)
\end{equation}

The Markov (short correlation time) approximation \cite%
{Klyatskin1991,Breuer2016,Hall2014,Tarasov2021} assumes that correlations $%
C_{j\ell }^{\prime }(t)$ in the definition of $A_{j\ell }^{\prime }(t)$
decay rapidly and that the upper limit in the correlation integral can be
taken to infinity. Then $A_{j\ell }^{\prime },$ $\gamma _{j\ell }$, $H_{%
\mathrm{eff}},$ $\Gamma _{nm}^{B}$, $\Lambda _{nm}$ and other related values
become constant, leading to a \emph{Lindblad equation }\cite%
{Lindblad1976,GoriniKossakowskiSudarshan1976,Breuer2007book,LindbladIntro2020}%
. In the pure-dephasing case, this approximation is generally sufficient;
the double sum over $j$ and $\ell $ may then be diagonalised so that the
dissipator is evaluated, as in (\ref{eq:rho}), over a single index $j$.

\subsection{Zurek's approach to dephasing\label{appBZ}}

This section presents an alternative approach to analysing decoherence that
appears more mathematically self-contained, but is more restrictive in the
context of environment-induced CP violations. The approach given by Zurek's
decoherence theory \cite{Zurek1982Environment} is based on the unperturbed
formulation and on the assumption that all coupling operators are diagonal
in the bath energy basis $|k\rangle $, \ $k=1,2,...$, and in the energy
basis of the surroundings $|\alpha \rangle $, \ $\alpha =1,2,...$, so that 
\begin{eqnarray}
V_{j} &=&\sum_{k}v_{jk}|k\rangle \langle k|,\ \ H_{B}|k\rangle
=E_{k}|k\rangle ,\text{ \ \ }\left[ H_{B},V_{j}\right] =0
\label{eq:Vj-zurek} \\
h_{BG} &=&\sum_{k\alpha }h_{k\alpha }|k\rangle \langle k|\otimes |\alpha
\rangle \langle \alpha |,\ \ \ H_{G}|\alpha \rangle =E_{\alpha }|\alpha
\rangle
\end{eqnarray}%
Tracing over the surroundings and the bath, $\rho _{S}(t)=\func{Tr}%
_{BG}\left( \rho (t)\right) $, after integrating the von Neumann equation (%
\ref{von-Neumann}) under the assumption of an initially factorised state $%
\rho (0)=\rho _{S}(0)\otimes \rho _{BG}(0)$ and using the diagonal product
basis defined above yields 
\begin{equation}
\rho _{nm}^{S}(t)\overset{\text{def}}{=}\left\langle n\right\vert \rho
_{S}(t)\left\vert m\right\rangle =\rho _{nm}^{S}(0)\exp \!\left( -\frac{i}{%
\hbar }(E_{n}-E_{m})t\right) D_{nm}(t),
\end{equation}%
where 
\begin{equation}
D_{nm}(t)=\sum_{k,\alpha }\rho _{k\alpha ,k\alpha }^{BG}(0)\Omega
_{nmk}(t)=\sum_{k}\rho _{kk}^{B}(0)\Omega _{nmk}(t)
\end{equation}%
is Zurek's pure decoherence factor satisfying $D_{nn}(t)=1.$ Here we define 
\begin{equation}
\Omega _{nmk}(t)\overset{\text{def}}{=}\exp \!\left( -\frac{it}{\hbar }%
\sum_{j}(\lambda _{jn}-\lambda _{jm})v_{jk}\right) ,\ \ \ \rho _{ki}^{B}(t)%
\overset{\text{def}}{=}\sum_{\alpha }\rho _{k\alpha ,i\alpha }^{BG}(t)
\end{equation}%
and note that $\Omega _{nmk}$ does not depend on $\alpha $ while $\rho
_{kk}^{B}=\rho _{kk}^{B}(0)$ remains steady-state and only diagonal elements 
$\rho _{kk}^{B}(0)$ affect $D_{nm}(t)$. The phase from $H_{BG}$ cancels when
the environment is traced out.

The diagonal assumptions also simplify the correlation $C_{j\ell }$ that
appears in the Redfield-Lindblad equation. Under these diagonal assumptions, 
$W_{j}^{(0)}(-\tau )=W_{j}(0),$ $W_{j}W_{\ell }=V_{j}V_{\ell }\otimes I_{G}$%
, $V_{j}V_{\ell }=\sum_{k}v_{jk}v_{\ell k}|k\rangle \langle k|$ and the
correlation matrix 
\begin{equation}
C_{j\ell }=\func{Tr}_{BG}\left( W_{j}W_{\ell }\rho _{BG}\right) =\func{Tr}%
_{B}\left( V_{j}V_{\ell }\rho _{B}\right) ,\ \ \ \ \rho _{B}\overset{\text{%
def}}{=}\func{Tr}_{G}\left( \rho _{BG}\right)
\end{equation}%
and its deviation $C_{j\ell }^{\prime }$ become time-independent.\ If the
supports in $k$-space of the matrices $V_{j}$ and $V_{\ell }$ do not
overlap, then the raw cross-correlations vanish, $C_{j\ell }=0$ when $j\neq
\ell $, while the centred correlations $C_{j\ell }^{\prime }$ generally need
not vanish. The assumption $v_{jk}=\delta _{jk}v_{k}$ leads to the further
simplification 
\begin{equation}
V_{k}=v_{k}|k\rangle \langle k|,\ \ \ \mu _{k}=v_{k}\langle k|\rho
_{B}|k\rangle ,\ \ \ C_{j\ell }=\delta _{j\ell }C_{j}
\end{equation}

The Markov approximation is based on the rapid decay of $C_{j\ell }^{\prime
}(t)$ as $t\rightarrow \infty $, which is generally not the case under
Zurek's diagonal approximation \cite%
{Zurek1982Environment,Zurek2003DecoherenceEinselection}. Hence, the Markov
path and the Zurek path represent alternative assumptions leading to
generally different treatments of dephasing.

\section{CP invariance of the system in the presence of environmental
interference \label{AppC}}

\subsection{CP-neutrality and CP-diagonalisation}

A system or environment is called \emph{CP-neutral} when each of its states $%
\left\vert k\right\rangle $ is mapped onto itself by the CP transformation 
\begin{equation}
\mathrm{U}_{\text{CP}}\left\vert k\right\rangle =+\left\vert k\right\rangle
\end{equation}%
leading to the following lemma

\begin{lemma}
\label{Lem2}Any CP-neutral system is always CP-balanced $\mathcal{U}_{\text{%
CP}}\left( \rho \right) \overset{\text{def}}{=}\bar{\rho}=\rho $ while any
operator $V$ acting in a neutral Hilbert space $\{\left\vert k\right\rangle
\}$ is CP-even $\mathcal{U}_{\text{CP}}\left( V\right) =V$.
\end{lemma}

Indeed applying CP conjugation to 
\begin{equation}
V=\sum_{k,i}V_{ki}\left\vert k\right\rangle \left\langle i\right\vert
\end{equation}
yields%
\begin{equation}
\mathcal{U}_{\text{CP}}\left( V\right) =\mathrm{U}_{\text{CP}}V\mathrm{U}_{%
\text{CP}}^{-1}=\sum_{k,i}V_{ki}\mathrm{U}_{\text{CP}}\left\vert
k\right\rangle \left\langle i\right\vert \mathrm{U}_{\text{CP}}^{\dagger
}=\sum_{k,i}V_{ki}\left\vert k\right\rangle \left\langle i\right\vert =V
\end{equation}

A system or environment which is not neutral can nevertheless be represented
in a diagonally neutral form. \ Consider $\left\vert k\right\rangle $ to be
bath energy eigenstates $H_{B}\left\vert k\right\rangle =E_{k}\left\vert
k\right\rangle .$\ Assume 
\begin{equation}
\mathrm{U}_{\text{CP}}\left\vert k\right\rangle =\left\vert \bar{k}%
\right\rangle ,\ \ \mathrm{U}_{\text{CP}}\left\vert \bar{k}\right\rangle
=\left\vert k\right\rangle ,\ \ \bar{k}\neq k
\end{equation}%
and define 
\begin{equation}
\left\vert k_{+}\right\rangle =\frac{\left\vert k\right\rangle +\left\vert 
\bar{k}\right\rangle }{\sqrt{2}},\ \ \ \left\vert k_{-}\right\rangle =\frac{%
\left\vert k\right\rangle -\left\vert \bar{k}\right\rangle }{\sqrt{2}}
\end{equation}%
hence $\mathrm{U}_{\text{CP}}\left\vert k_{+}\right\rangle =+\left\vert
k_{+}\right\rangle $ \ and \ $\mathrm{U}_{\text{CP}}\left\vert
k_{-}\right\rangle =-\left\vert k_{-}\right\rangle $ and, since $%
H_{B}\left\vert \bar{k}\right\rangle =E_{\bar{k}}\left\vert \bar{k}%
\right\rangle $\ with $E_{\bar{k}}=E_{k}$ (degeneracy that follows from
CP-invariance of $H_{B}$, since $H_{B}\left\vert \bar{k}\right\rangle =H_{B}%
\mathrm{U}_{\text{CP}}\left\vert k\right\rangle =\mathrm{U}_{\text{CP}%
}H_{B}\left\vert k\right\rangle =E_{k}\left\vert \bar{k}\right\rangle $), $%
H_{B}\left\vert k_{+}\right\rangle =E_{k}\left\vert k_{+}\right\rangle $ and 
$H_{B}\left\vert k_{-}\right\rangle =E_{k}\left\vert k_{-}\right\rangle $.
Therefore, in the new energy eigenstate basis $\left\vert i\right\rangle
=\left\{ \left\vert k_{+}\right\rangle ,\left\vert k_{-}\right\rangle
,...\right\} $ we have%
\begin{equation}
\mathrm{U}_{\text{CP}}\left\vert i\right\rangle =\xi _{i}\left\vert
i\right\rangle ,\ \ \ \text{where }\xi _{i}=\pm 1
\end{equation}%
and therefore%
\begin{equation}
\mathcal{U}_{\text{CP}}\left( V\right) =\sum_{i,\ell }V_{i\ell }\mathrm{U}_{%
\text{CP}}\left\vert i\right\rangle \left\langle \ell \right\vert \mathrm{U}%
_{\text{CP}}^{\dagger }=\sum_{i,\ell }V_{i\ell }\xi _{i}\xi _{\ell }^{\ast
}\left\vert i\right\rangle \left\langle \ell \right\vert \neq V
\end{equation}%
If all $\xi _{i}=1$, the system is CP-neutral; if all $\xi _{i}=-1$, the
system can be transformed into CP-neutral by defining new $\mathrm{U}_{\text{%
CP}}$ as $-\mathrm{U}_{\text{CP}}$. A system with alternating signs of $\xi
_{i}$ may not be CP-balanced since $\xi _{i}\xi _{\ell }^{\ast }=\pm 1.$
This representation, which is chosen to consist of energy eigenstates, is
called \emph{CP-diagonalised.}

\begin{lemma}
\label{Lem3}In CP-diagonalised basis, the diagonal projection $V^{\text{diag}%
}$ of any operator $V$ is preserved by CP transformation $\mathcal{U}_{\text{%
CP}}(V^{\text{diag}})=V^{\text{diag}}$ (i.e. is CP-even). The same statement
applies to the diagonal projection of the density $\mathcal{U}_{\text{CP}%
}(\rho ^{\text{diag}})=\rho ^{\text{diag}}.$\ 
\end{lemma}

Indeed, the diagonal operator 
\begin{equation}
V^{\text{diag}}=\sum_{i}V_{ii}\left\vert i\right\rangle \left\langle
i\right\vert
\end{equation}%
is transformed under CP as 
\begin{equation}
\mathcal{U}_{\text{CP}}\left( V^{\text{diag}}\right) =\sum_{i}V_{ii}\mathrm{U%
}_{\text{CP}}\left\vert i\right\rangle \left\langle i\right\vert \mathrm{U}_{%
\text{CP}}^{\dagger }=\sum_{i}V_{ii}\xi _{i}\xi _{i}^{\ast }\left\vert
i\right\rangle \left\langle i\right\vert =+V^{\text{diag}}
\end{equation}%
since $\xi _{i}\xi _{i}^{\ast }=+1.$

\subsection{Fundamental CP invariance of Redfield equations}

Consider the CP decomposition for $h_{SB}$ expansion in (\ref{eq:hSB}) 
\begin{eqnarray}
L_{j+} &=&\frac{L_{j}+\mathcal{U}_{\text{CP}}(L_{j})}{2},\ \ \ V_{j+}=\frac{%
V_{j}+\mathcal{U}_{\text{CP}}(V_{j})}{2}  \label{eq:CP-decomp} \\
L_{j-} &=&\frac{L_{j}-\mathcal{U}_{\text{CP}}(L_{j})}{2},\ \ \ V_{j-}=\frac{%
V_{j}-\mathcal{U}_{\text{CP}}(V_{j})}{2}
\end{eqnarray}%
Taking the sum over all $j$ gives 
\begin{equation}
h_{SB}=\sum_{j}\left( L_{j+}\otimes V_{j+}+L_{j-}\otimes V_{j-}\right)
\label{h_SB}
\end{equation}%
where the odd cross-products $L_{j+}\otimes V_{j-}$ and $L_{j-}\otimes
V_{j+} $ are omitted from the CP-invariant representation since these terms
must cancel when sum over $j$ is evaluated or, otherwise, the principal
assumption that $h_{SB}$ is CP-even 
\begin{equation}
\mathcal{U}_{\text{CP}}(h_{SB})=+h_{SB}
\end{equation}%
would be violated. As $L_{j+}$ and $V_{j+}$ are CP-even 
\begin{equation}
\mathcal{U}_{\text{CP}}(L_{j+})=+L_{j+},\ \ \mathcal{U}_{\text{CP}%
}(V_{j+})=+V_{j+}
\end{equation}%
and $L_{j-}$ and $V_{j-}$ are CP-odd 
\begin{equation}
\mathcal{U}_{\text{CP}}(L_{j-})=-L_{j-},\ \ \mathcal{U}_{\text{CP}%
}(V_{j-})=-V_{j-}
\end{equation}%
the system-bath interaction Hamiltonian remains CP-invariant (i.e. CP-even) $%
\mathcal{U}_{\text{CP}}(h_{SB})=+h_{SB}$. Indeed, $\mathcal{U}_{\text{CP}%
}(L_{j-}\otimes V_{j-})=\mathcal{U}_{\text{CP}}(L_{j-})\mathcal{U}_{\text{CP}%
}(V_{j-})=+L_{j-}\otimes V_{j-}$.

This yields the following lemma.

\begin{lemma}
\label{lemA1}The Redfield operator remains CP-even when:

\begin{enumerate}
\item no CP-odd terms $L_{j-}$ are present in the expansion

\item the CP-odd terms $L_{j-}$ are present but the coefficients $\mu _{j-},$
$C_{j-\ell +}^{\prime }(\tau )$ and $C_{\ell +j-}^{\prime }(\tau )$
multiplying the corresponding CP-odd operators $L_{j-}$, $L_{j-}L_{\ell +}$
and $L_{\ell +}L_{j-}$ are either zero or change sign under the
transformation.
\end{enumerate}
\end{lemma}

The transformation of the coefficients is covered by the following lemma.

\begin{lemma}
\label{lemA2}Under the fundamental CP transformation $\mathcal{U}_{\text{CP}%
}^{\text{full}}(...)$, the following coefficients of the Redfield equation, $%
\mu _{j-},$ $C_{j-\ell +}^{\prime }(\tau )$ and $C_{\ell +j-}^{\prime }(\tau
)$, change sign, while $\mu _{j+},$ $C_{j+\ell +}^{\prime }(\tau )$ and $%
C_{\ell -j-}^{\prime }(\tau )$ keep their signs.
\end{lemma}

Indeed, 
\begin{eqnarray}
\mu _{j\pm }(\rho _{B}) &=&\func{Tr}_{B}(V_{j\pm }\rho _{B})=\func{Tr}%
_{B}\left( \mathcal{U}_{\text{CP}}(V_{j\pm }\rho _{B})\right) \\
&=&\pm \func{Tr}_{B}\left( V_{j\pm }\bar{\rho}_{B}\right) =\pm \mu _{j\pm }(%
\bar{\rho}_{B})  \notag
\end{eqnarray}%
\begin{eqnarray}
C_{j-\ell +}(\tau ;\rho _{BG}) &=&\func{Tr}_{BG}\left( W_{j-}W_{\ell
+}^{(0)}(-\tau )\rho _{BG}\right)  \notag \\
&=&\func{Tr}_{BG}\left( \mathcal{U}_{\text{CP}}\left( W_{j-}W_{\ell
+}^{(0)}(-\tau )\rho _{BG}\right) \right) \\
&=&-\func{Tr}_{BG}\left( W_{j-}W_{\ell +}^{(0)}(-\tau )\bar{\rho}%
_{BG}\right) =-C_{j-\ell +}(\tau ;\bar{\rho}_{BG})  \notag
\end{eqnarray}%
and 
\begin{equation}
C_{j-\ell +}^{\prime }(\tau ;\rho _{BG})=-C_{j-\ell +}^{\prime }(\tau ;\bar{%
\rho}_{BG})
\end{equation}%
since $C_{j-\ell +}^{\prime }=C_{j-\ell +}-\mu _{j-}\mu _{\ell +}.$ A
similar argument applies to $C_{\ell +j-}^{\prime }$ and the other
coefficients.

If the overall Hamiltonian is CP-invariant $\mathcal{U}_{\text{CP}}(H)=H$
(i.e. no fundamental CP violations are present\ \cite%
{Symmetry1972,PDGConservation2020,PDG2024_RPP}), then the von Neumann
equation (\ref{von-Neumann}) is CP-covariant (\ref{von-N-cov}). Since the
Redfield equations (\ref{red1})-(\ref{rede2}) are approximate and,
therefore, may or may not preserve exact CP invariance, this requires
special analysis. The outcome is given by the following theorem.

\begin{theorem}
\label{theorA1}The Redfield dephasing equations preserve fundamental CP
invariance of the underlying Hamiltonian.
\end{theorem}

The \emph{fundamental CP covariance} of the Redfield equations (\ref{red1})-(%
\ref{rede2}) is specified by (\ref{Red-inv0}) and immediately follows from
lemmas \ref{lemA1} and \ref{lemA2} under conditions when the underlying
overall Hamiltonian is CP-invariant $\mathcal{U}_{\text{CP}}(H)=H$.

\subsection{Apparent CP invariance}

We now turn to apparent CP transformation, which is applied to the system
but not to the environment---see (\ref{fund-app}). This transformation is
denoted here as $\mathcal{U}_{\text{CP}}^{\text{app}}(...)$ \ Although this
work assumes that all Hamiltonians are CP-invariant, the possibility of
apparent CP bias is underpinned by the possible presence of CP-odd
coefficients in the Redfield operators, and \emph{fundamental CP invariance}
(\ref{Red-inv0}) does not necessarily transfer into \emph{apparent CP
invariance} (\ref{Red-inv1}). However, apparent CP invariance is preserved
under a number of conditions listed below.

\begin{lemma}
\label{lemC1} When the bath is CP-balanced, $\bar{\rho}_{B}=\rho _{B}$, the
system remains apparently CP-invariant at the first order (i.e. $\mathcal{R}%
^{(1)}$), but generally not at the second order (i.e $\mathcal{R}^{(2)}$).
\end{lemma}

According to lemma \ref{lemA2}, 
\begin{equation}
\mu _{j-}(\rho _{B})=-\mu _{j-}(\bar{\rho}_{B})=-\mu _{j-}(\rho _{B})=0
\end{equation}%
when $\bar{\rho}_{B}=\rho _{B}$, and the first-order term $\mathcal{R}_{\rho
_{BG}}^{(1)}$ in (\ref{red1}) has $\mu _{j-}=0$. Note that, generally, $%
C_{j\ell }^{\prime }(\tau ;\rho _{BG})\neq 0$\ \ due to the influence of
unbalanced surroundings on coefficients.

\begin{lemma}
\label{lemC2}The system remains apparently CP-invariant when the environment
consisting of the bath and surroundings is CP-balanced, $\bar{\rho}%
_{BG}=\rho _{BG}$.
\end{lemma}

According to lemma \ref{lemA2}, 
\begin{equation}
C_{j+\ell -}^{\prime }(\tau ;\rho _{BG})=-C_{j+\ell -}^{\prime }(\tau ;\bar{%
\rho}_{BG})=-C_{j+\ell -}^{\prime }(\tau ;\rho _{BG})=0
\end{equation}%
when $\bar{\rho}_{BG}=\rho _{BG}$, and all coefficients multiplying CP-odd
operator structures vanish in (\ref{red1}).

\begin{lemma}
\label{lemC3}The system remains apparently CP-invariant when the bath is
CP-neutral.
\end{lemma}

According to lemma \ref{Lem2}, all operators $V_{j}$ acting on a CP-neutral
bath are CP-even. Hence, according to (\ref{h_SB}), all operators $L_{j}$
are also CP-even.

\begin{lemma}
\label{LemC4}The system remains apparently CP-invariant when the interaction
Hamiltonian $h_{SB}$ is diagonal in a CP-diagonalised basis.
\end{lemma}

According to lemma \ref{Lem3}, all operators $V_{j}$ that are diagonal in a
CP-diagonalised basis are CP-even. Hence, according to (\ref{h_SB}), all
operators $L_{j}$ are also CP-even.

Under conditions specified by these lemmas, the bath becomes \emph{%
CP-screening} and the system remains apparently CP-invariant.

\nocite{PDG2020_RPP,PDG2022_RPP} 
\bibliographystyle{amsalpha}
\bibliography{HEP}

\end{document}